\documentclass[namedreferences]{SolarPhysics}
\usepackage[optionalrh]{spr-sola-addons}
\usepackage{graphicx}        
\usepackage{color}           
\usepackage{url}             

\newcommand{\etal}{{\it et al.}}

\newcommand{\aap}{    {\it Astron. Astrophys.}}

\newcommand{\apj}{    {\it Astrophys. J.}}

\newcommand{\jgr}{    {\it J. Geophys. Res.}}

\newcommand{\solphys}{{\it Solar Phys.}}

\begin{document}

\begin{article}

\begin{opening}

\title{3D MHD Coronal Oscillations About a Magnetic Null Point: Application of WKB Theory}

\author{J.A. \surname{McLaughlin} \sep  J.S.L. \surname{Ferguson} \sep A.W. \surname{Hood}}
\runningauthor{J.A. McLaughlin {\it{et al.}}}
\runningtitle{Application of WKB Theory to 3D MHD}

   \institute{School of Mathematics and Statistics, University of St Andrews, St Andrews, Fife, KY16 9SS, UK
                     email: \url{james@mcs.st-and.ac.uk}\\ 
             }

\date{Received 27 September 2007; accepted 4 December 2007}

\begin{abstract}
{ This paper is a demonstration of how the WKB approximation can be used to help solve the linearised 3D MHD equations. Using Charpit's Method and a Runge-Kutta numerical scheme, we have demonstrated this technique for a potential 3D magnetic null point, ${\bf{B}}=\left( x,\epsilon y -\left(\epsilon +1\right)z\right)$. Under our cold plasma assumption, we have considered two types of wave propagation: fast magnetoacoustic and Alfv\'en waves. We find that the  fast magnetoacoustic wave experiences refraction towards the magnetic null point, and that the effect of this refraction depends upon the Alfv\'en speed profile.  The wave, and thus the wave energy, accumulates at the null point. We have found that current build up is exponential and the exponent is dependent upon $\epsilon$.  Thus, for the fast wave there is preferential heating at the null point. For the Alfv\'en wave, we find that the wave propagates along the fieldlines.  For an Alfv\'en wave generated along the fan-plane, the wave accumulates along the spine. For an Alfv\'en wave generated across the spine,  the value of $\epsilon$ determines where the wave accumulation will occur:  fan-plane ($\epsilon=1$), along the $x-$axis ($0<\epsilon <1$) or along the $y-$axis ($\epsilon>1$). We have shown analytically that currents build up  exponentially, leading to preferential heating in these areas. The work described here highlights the importance of understanding the magnetic topology of the coronal magnetic field for the location of wave heating.}
\end{abstract}
\end{opening}

\section{Introduction}

The WKB approximation is an asymptotic approximation technique which can be used when a system contains a large parameter (see {\it{e.g.}} \opencite{Bender}). Hence, the WKB method can be used in a system where a wave propagates through a background medium which varies on some spatial scale which is much longer than the wavelength of the wave. The SOHO and TRACE satellites have recently observed MHD wave motions in the corona, {\it{i.e.}} fast and slow magnetoacoustic waves and Alfv\'en waves (see reviews by \opencite{NV2005}; \opencite{Ineke2005}; \citeyear{Ineke2006}).  The coronal magnetic field  plays a fundamental role in their propagation and to begin to understand this  inhomogeneous magnetised  environment, it is useful to look at the  structure (topology) of the magnetic field itself.  Potential-field extrapolations of the coronal magnetic field can be made from photospheric magnetograms. Such extrapolations show the existence of an important feature of the topology: {\it{null points}}. Null points are points in the field where the magnetic field, and hence the Alfv\'en speed, is zero. Detailed investigations of the coronal magnetic field, using such potential field calculations, can be found in \inlinecite{Beveridge2002} and \inlinecite{Brown2001}.

\inlinecite{MH2004} found that for a single 2D null point, the fast magnetoacoustic wave was attracted to the null and the wave energy accumulated there. In addition, they found that the Alfv\'en wave energy accumulated along the separatrices of the topology. They solved the 2D linearised MHD equations numerically and compared the results with a WKB approximation: the agreement was excellent. From their work and other examples ({\it{e.g.}} \opencite{Galsgaard2003}; \opencite{MH2005}; \citeyear{MH2006}; \opencite{Khomenko2006}) it has been clearly demonstrated that  the WKB approximation can provide a vital link between analytical and numerical work, and often provides the critical insight to understanding the physical results. This paper demonstrates the methodology of  how to apply the WKB approximation in linear 3D MHD. We believe that with the vast amount of 3D modelling currently being undertaken, applying this WKB technique to 3D will be very useful and beneficial to modellers in the near future.

The work undertaken by \inlinecite{Galsgaard2003} deserves special mention here. They performed numerical experiments on the effect of twisting the spine of a 3D null point, and described the resultant wave propagation towards the null. They found that when the fieldlines around the spine are perturbed in a rotationally symmetric manner, a twist wave (essentially an Alfv\'en wave) propagates towards the null along the fieldlines.  Whilst this Alfv\'en wave spreads out as the null is approached, a fast-mode wave focuses on the null and wraps around it. They concluded that the driving of the fast wave was likely to come from a non-linear coupling to the Alfv\'en wave \cite{Nak1997}. They also compare their results with a WKB approximation and find that, for the $\beta=0$ fast wave, the wavefront wraps around the null point as it contracts towards it. They perform their WKB approximation in cylindrical polar coordinates and thus their resultant equations are two-dimensional (since a simple 3D null point is essentially 2D in cylindrical coordinates). In contrast, we solve the WKB equations for three Cartesian components, and thus we can solve for more general disturbances and more general boundary conditions. This also allows us to concentrate on the transient features that are not always apparent when  only cylindrically symmetric solutions are permitted. 

More recently, \inlinecite{PG2007} and \inlinecite{PBG2007} have performed numerical simulations in which the spine and fan of a 3D null point are subject to rotational and shear perturbations. They found that  rotations of the fan plane lead to current sheets in the location of the spine and rotations about the spine lead to current sheets  in the fan. In addition,  shearing perturbations lead to 3D localised current sheets focused at the null point itself.  This general behaviour is in good agreement with the work presented in this paper, {\it{i.e.}} current accumulation at certain parts of the topology. However,  the primary motivation in \inlinecite{PG2007} and \inlinecite{PBG2007} was to investigate current-sheet formation and reconnection rates, whereas the techniques described in this paper focus on MHD wave-mode propagation and interpretation.

The propagation of fast magnetoacoustic waves in an inhomogeneous coronal plasma has been investigated by \inlinecite{NR1995}, who showed how the waves are refracted into regions of low Alfv\'en speed. In the case of null points, the Alfv\'en speed actually drops to zero.

The paper has the following outline: In Section \ref{SEC:1}, the basic equations are described. Section \ref{WKBAPPROXIMATION} details the 3D WKB approximation utilised in this paper. The results for the fast wave and Alfv\'en waves are shown in Section \ref{SEC:FAST} and \ref{SEC:ALFVEN}. The conclusions and discussion are presented in Section \ref{conclusion}. There are four appendices which complement the results in the main text.


\section{Basic Equations}\label{SEC:1}

The usual resistive, adiabatic MHD equations for a plasma in the solar corona are used:
\begin{eqnarray}
\rho{\partial {\bf v} \over \partial t} + \rho \left ({\bf v}\cdot \nabla \right){\bf v} &=& - \nabla p +{\bf j} \times {\bf B} + \rho {\bf{g}}  \label{1} \;\;,\\
{\partial {\bf B} \over \partial t} &=& \nabla \times \left ({\bf v} \times {\bf B} \right) + \eta \nabla^{2} {\bf B}\;\;,\label{2}\\
{\partial {\rho} \over \partial t} + \nabla \cdot \left (\rho{\bf v} \right) &=& 0\;\;,\label{3}\\
{\partial p \over \partial t} + {\bf v} \cdot \nabla p &=& - \gamma p \nabla. {\bf v}\;\;,\label{3b}\\
\mu \:{\bf{j}} &=& \nabla \times {\bf{B}}\;\; ,\label{4}
\end{eqnarray}
where ${\bf v}$  is the plasma velocity, ${\rho}$  is the mass density, ${p}$  is the gas pressure, ${\bf B}$  is the magnetic induction (usually called the magnetic field), ${\bf{j}}$ is the electric current, ${\bf{g}}$ is gravitational acceleration, ${\gamma}$  is the ratio of specified heats, ${\eta}$  is the magnetic diffusivity and $\mu$ is the magnetic permeability.


\subsection{Basic Equilibrium}\label{mageqn}

\begin{center}
\begin{figure}[t]
\hspace{-1.0cm}
\includegraphics[scale=0.38]{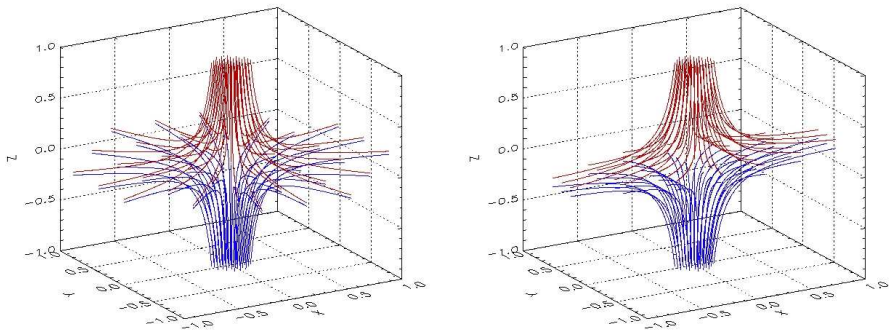}
\caption{{\emph{Left}}: Proper radial null point, described by ${\bf{B}}=(x,y,-2z)$, {\it{i.e.}} $\epsilon=1$. {\emph{Right}}: Improper radial null point, described by ${\bf{B}}=(x,\frac{1}{2}y,-\frac{3}{2}z)$, {\it{i.e.}} $\epsilon=\frac{1}{2}$.  Note for $\epsilon=\frac{1}{2}$, the field lines rapidly curve such that they run parallel to the $x-$axis along $y=0$. In both figures, the  $z-$axis indicates the {\emph{spine}} and the $xy-$plane at $z=0$ denotes the {\emph{fan}}. The red fieldlines have been tracked from the $z=1$ plane, the blue from $z=-1$.}
\label{figure1}
\end{figure}
\end{center}

We choose a  3D magnetic null point for our equilibrium field, of the form:
\begin{eqnarray}
{\bf{B}}_0 = \frac{B}{L} \left(x,\epsilon y,-\left( \epsilon +1 \right) z \right)   \label{Bfield}  \;\;,
\end{eqnarray}
where $B$ is a characteristic field strength, $L$ is the length scale for magnetic field variations and the parameter $\epsilon$ is related to the predominate direction of alignment of the fieldlines in the fan plane.  \inlinecite{Parnell1996} investigated and classified the different types of linear magnetic null points that can exist (our $\epsilon$ parameter is called $p$ in their work). Topologically, this 3D null consists of two key parts: the $z-$axis represents a special, isolated fieldline called the {\emph{spine}} which approaches the null from above and below \cite{PriestTitov1996} and the $xy-$plane through $z=0$ is known as the {\emph{fan}} and consists of a surface of fieldlines spreading out radially from the null. Figure \ref{figure1} shows two examples of 3D null points: $\epsilon=1$ (left) and  $\epsilon=1/2$ (right). \inlinecite{TH2000} have investigated the steady state structures of magnetic null points.

Equation (\ref{Bfield}) is the general expression for the linear field about a potential magnetic null point (\opencite{Parnell1996}: Section IV).  In this paper, we only consider $\epsilon \ge 0$ and so all nulls we describe are {\it{positive}} nulls, {\it{i.e.}} the spine points into the null and the field lines in the fan are directed away. In addition, all potential nulls are designated  {\it{radial}}, {\it{i.e.}} there is no spiral motions in the  fan-plane. In general, there are three cases to consider:
\begin{itemize}
\item{{\bf{$\epsilon  = 1$}}: describes a {\it{proper null}} (Figure \ref{figure1}: Left). This magnetic null has cylindrical symmetry about the spine axis.}
\item{{\bf{$\epsilon > 0,\: \epsilon \neq 1$}}: describes an {\it{improper null}}  (Figure \ref{figure1}: Right). Field lines rapidly curve such that they run parallel to the $x-$axis if $0<\epsilon <1$ and parallel to the $y-$axis if $\epsilon >1$.}
\item{{\bf{$\epsilon=0$}}: equation (\ref{Bfield}) reduces to the X-point potential field in the $xz-$plane and forms a null line along the $y-$axis through $x=z=0$. MHD wave propagation in this 2D configuration has been studied extensively by \citeauthor{MH2004} (\citeyear{MH2004}; \citeyear{MH2005}; \citeyear{MH2006}).}
\end{itemize} 



\subsection{Assumptions and Simplifications}

In this paper, the linearised MHD equations are used to study the nature of wave propagation near the null point. Using subscripts of $0$ for equilibrium quantities and $1$ for perturbed quantities, Equations (\ref{1})\,--\,(\ref{4}) become:
\begin{eqnarray}
\rho_{0}{\partial {\bf v}_{1} \over \partial t} &=& - \nabla p_{1} +{\bf {j}}_{0}  \times {{\bf B}}_1 + {{\bf {j}}}_1 \times {\bf B}_{0}  + \rho_1 {\bf{g}}\;\;, \label{5}\\
{\partial {{\bf B}}_1 \over \partial t} &=& \nabla \times \left ({\bf v}_{1} \times {\bf B}_{0} \right) + \eta \nabla^{2} {{\bf B}}_1   \;\;,    \label{6}\\
{\partial {\rho_{1}} \over \partial t} &+& \nabla \cdot \left (\rho_{0}{\bf v}_{1} \right) = 0     \;\;,    \label{7}\\
{\partial p_{1} \over \partial t} &+& {\bf v}_{1}.\nabla p_{0} = - \gamma p_{0} \nabla \cdot {\bf v}_{1}   \;\;,        \label{7b}\\
\mu \:{\bf{j}}_1 &=& \nabla \times {\bf{B}}_1\;\; . \label{8}
\end{eqnarray}

We now consider several simplifications to our system. We will only be  considering a potential equilibrium magnetic field ($\nabla \times  {\bf B}_{0}={\bf{0}}$) in an ideal system ($\eta=0$). We will also assume the equilibrium gas density ($\rho_0$) is uniform. A spatial variation in $\rho _0$ can cause phase mixing \cite{Heyvaerts1983,DeMoortel1999,Hood2002}. In addition, we ignore the effect of gravity on the system ({\it{i.e.}} we set $ {\bf{g}} = 0$). Finally, in this paper we assume a cold plasma, {\it{i.e.}} $c_s=\sqrt{\gamma p_0 / \rho_0}=0$.

We will not discuss Equation (\ref{7}) further as it can be solved once we know $\mathbf{v} _1$. In fact, under the assumptions of linearisation and no gravity, it has no influence on the momentum equation  and so in effect the plasma is arbitrarily compressible \cite{CW1992}.


We now non-dimensionalise the above equations as follows: let ${\mathbf{\mathrm{v}}}_1 = \bar{\rm{v}} {\mathbf{v}}_1^*$, ${\mathbf{B}}_0 = B {\mathbf{B}}_0^*$, ${\mathbf{B}}_1 = B{\mathbf{B}}_1^*$,    $x = L x^*$, $z=Lz^*$, $\nabla = \frac{1}{L}\nabla^*$ and $t=\bar{t}t^*$, where we let $*$ denote a dimensionless quantity and $\bar{\rm{v}}$, $B$,       $L$, and $\bar{t}$ are constants with the dimensions of the variable that they are scaling. In addition, $\rho_0$ and $p_0$ are constants as these equilibrium quantities are uniform ({\it{i.e.}} $\rho_0^*=p_0^*=1$). We then set $ {B} / {\sqrt{\mu \rho _0 } } =\bar{\rm{v}}$ and $\bar{\rm{v}} =  {L} / {\bar{t}}$ (setting $\bar{\rm{v}}$ as a constant background Alfv\'{e}n speed). Under these scalings, $t^*=1$ (for example) refers to $t=\bar{t}=  {L} / {\bar{\rm{v}}}$; {\it{i.e.}} the (background) Alfv\'en time taken to travel a distance $L$.  For the rest of this paper, we drop the star indices; the fact that they are now non-dimensionalised is understood.

These non-dimensionalised equations can be combined to form one single equation:
\begin{eqnarray}
\frac{\partial^2 } {\partial t^2}{\mathbf{v}}_1 =  \left\{ \nabla \times \left[  \nabla \times \left( {\mathbf{v}}_1 \times {\mathbf{B}}_0 \right) \right] \right\} \times {\mathbf{B}}_0   \;\;. \label{10}
\end{eqnarray}


\section{WKB Approximation}\label{WKBAPPROXIMATION}

In this paper, we will be looking for WKB solutions of the form:
\begin{eqnarray}
\mathbf{v} = {\bf{a}} {\rm{e}}^{{\rm{i}} \phi (x,y,z,t) }  \label{WKB}
\end{eqnarray}
where $\bf{a}$ is a constant. In addition, we define $\omega = \phi_t$ as the frequency and ${\bf{k}} = \nabla \phi=\left( \phi_x, \phi_y, \phi_z\right)=\left(p,q,r\right)$ as the wavevector. $\phi$, and its  derivatives, are considered to be the large parameters in our system.


One of the difficulties associated with 3D MHD wave propagation is distinguishing between the three different wave types, {\it{i.e.}} between the fast and slow magnetoacoustic waves and the Alfv\'en wave. To aid us in our interpretation, we now define a new coordinate system: $( {\bf B}_{0},  {\bf k},  {\bf B}_{0} \times {\bf k}$), where ${\bf{k}}$ is our wavevector as defined above. This coordinate system fully describes all three directions in space when ${\bf B}_{0}$ and  ${\bf {k}}$ are not parallel to each other, {\it{i.e.}} $                       {\bf {k}} \neq \lambda       {\bf B}_{0} $, where $\lambda$ is some constant of proportionality. In the work below, we will proceed assuming  $  {\bf {k}} \neq \lambda       {\bf B}_{0}        $. The scenario where  $   {\bf {k}} = \lambda       {\bf B}_{0}          $ is looked at in Appendix \ref{appendixA}. In fact, the work described below is also valid for  ${\bf k} = \lambda{\bf {B}}_0$ with the consequence that the solution is degenerate, {\it{i.e.}} the waves recovered are identical and cannot be distinguished.

We now substitute $\mathbf{v} = {\bf{a}} {\rm{e}}^{{\rm{i}} \phi (x,y,z,t) }$  into Equation (\ref{10}) and make  the WKB approximation such that $\phi \gg 1$. Taking the dot product with ${\bf{B}}_0$, $\bf{k}$ and ${\bf{B}}_0 \times {\bf{k}}$  gives three velocity components which, in matrix form, are:
\begin{eqnarray*}
\left[ \begin{array}{ccc}
\omega^2 & 0  & 0 \\
\left( { {\bf{B}}_0 \cdot  {\bf{k}} } \right) \left|    {\bf{k}}   \right|^2  & \; \; \omega^2 -  \left|    {\bf{B}}_0   \right|^2       \left|    {\bf{k}}    \right|^2  & 0 \\
0 & 0 &  \omega^2  - {\left( {\bf{B}}_0 \cdot   {\bf{k}}   \right)^2}\end{array}\right]
 \left( \begin{array}{c}
{\bf v}\cdot  {\bf B}_{0} \\
{\bf v}\cdot    {\bf{k}}  \\
{\bf v}\cdot {  {\bf B}_{0}\times{    {\bf{k}}    }}\end{array}\right)
=
 \left (\begin{array}{c}
0 \\
0 \\
0 \end{array}\right)
\end{eqnarray*}
The matrix of these three coupled Equations must have zero determinant so as not to have a trivial solution. Thus, taking the determinant gives:
\begin{eqnarray}
 \mathcal{F}\left( \phi, \omega, t,  {\bf{B}}_0, {\bf{k}} \right)=\left(\omega^{2} - 0\right)\left( \omega^{2} - {\left|{\bf B}_0 \right|^2 \left|    {\bf{k}} \right|^2 } \right)  \left( \omega^{2}-\left( {\bf{B}}_0 \cdot {\bf{k}} \right) ^{2} \right) =0 \;\;, \label{F1}
\end{eqnarray}
where $\mathcal{F}$ is a first-order, non-linear PDE. Equation (\ref{F1}) has two solutions, corresponding to two different MHD wave types (in general three, but the slow wave has vanished under the cold plasma approximation). The two solutions correspond to the fast magnetoacoustic wave and to the Alfv\'en wave.

In Sections \ref{SEC:FAST} and \ref{SEC:ALFVEN}, we will examine each of these wave solutions in detail for the 3D magnetic null point configuration described by Equation (\ref{Bfield}) for both $\epsilon=1/2$ and $\epsilon=1$. However, it should be noted that the technique described above is valid for any 3D magnetic configuration. The case where the two roots of Equation (\ref{F1}) are the same is examined in Appendix \ref{appendixA}.


\section{Fast Wave}\label{SEC:FAST}

Let us first consider the fast wave solution, and hence we assume $\omega^2 \neq \left( {\bf{B}}_0 \cdot {\bf{k}} \right) ^{2}$. Thus, Equation (\ref{F1}) simplifies to:
\begin{eqnarray}
{\mathcal{F}} \left( \phi, \omega, t, {\bf{B}}_0,  {\bf{k}} \right) &=& \omega^{2} -  \left| {\bf{B}}_0   \right| ^2  \left|   {\bf{k}} \right| ^2   \nonumber \\
&=&  \omega^{2} - \left(x^2+\epsilon^2 y^2+ \left( \epsilon+1\right)^2 z^2\right) \left( p^2+q^2+r^2\right)=0  \;\; \label{fastEquation}
\end{eqnarray}
We can now use Charpit's Method (see e.g. \opencite{Evans1999}) to solve this first-order PDE, where we assume our variables depend upon some independent parameter $s$ in characteristic space.  Charpit's Method replaces a first-order PDE with a set of characteristics that are a system of ODEs. Charpit's Equations take the form:
\begin{eqnarray*}
\frac{ {\rm{d}} \phi}{{\rm{d}} s} &=& \left( {{\omega}}\frac{\partial} {\partial {{\omega}}}    +    {\bf{k}}\cdot \frac{\partial} {\partial {\bf{k}}} \right) \mathcal{F}\;\;, \quad\frac{{\rm{d}} {{t}}} {{\rm{d}} s} =\frac{\partial} {\partial {{\omega}}} \mathcal{F}  \;\;,\quad \frac{{\rm{d}} {\bf{x}}} {{\rm{d}} s} =\frac{\partial} {\partial {\bf{k}}} \mathcal{F}     \;\;,\\
 \frac{{\rm{d}} {{\omega}}}{{\rm{d}} s}&=& -\left( \frac{\partial} {\partial {{t}}} + {{\omega}} \frac{\partial} {\partial \phi}  \right) \mathcal{F}\;\;,\quad  \frac{{\rm{d}} {\bf{k}}}{{\rm{d}} s}= -\left( \frac{\partial} {\partial {\bf{x}}} + {\bf{k}} \frac{\partial} {\partial \phi}  \right) \mathcal{F}\;\;,
\end{eqnarray*}
where, as previously defined, ${\bf{k}}=(p,q,r)$ and ${\bf{x}}=(x,y,z)$. In general, the coupled Equations have to be solved numerically, but analytical solutions have been found in 2D \cite{MH2004}. These ODEs are subject to the initial conditions $\phi=\phi_0(s=0)$, $x=x_0(s=0)$,  $y=y_0(s=0)$, $z=z_0(s=0)$,   $t=t_0(s=0)$, $p=p_0(s=0)$, $q=q_0(s=0)$, $r=r_0(s=0)$, and $\omega=\omega_0(s=0)$    and, in the following work, are solved numerically using a fourth-order Runge-Kutta method.

In addition, note that there are no boundary conditions in the usual sense: the variables are solved using Charpit's Method (essentially the method of characteristics) and the resulting characteristics are only dependent upon initial position $\left( x_0, y_0, z_0, t_0 \right)$ and distance travelled along the characteristic; $s$. Thus, there are no computational boundaries and no boundary conditions (only initial conditions). In this paper, we have chosen to illustrate our results in the domain $-1 \le x \le 1$, $-1 \le y \le 1$, $-1 \le z \le 1$, and this choice is arbitrary. That the WKB solutions are independent of  boundary conditions is actually an advantage over traditional numerical simulations; where the choice of boundary conditions can play a significant role.

For the fast wave (Equation \ref{fastEquation}) Charpit's Equations are:
\begin{eqnarray}
{{\rm{d}}\phi \over {\rm{d}}s}&=&0 \;\;, \quad{{\rm{d}}t \over {\rm{d}}s}=\omega \;\;,\quad {{\rm{d}}x \over {\rm{d}}s}=-pA \;\;,\quad{{\rm{d}}y \over {\rm{d}}s}=-qA \;\;,\quad{{\rm{d}}z \over {\rm{d}}s}=-rA  \nonumber\\
{{\rm{d}} \omega \over {\rm{d}}s}&=&0 \;\;,    \quad{{\rm{d}}p \over {\rm{d}}s}=\;xB \;\;,\quad{{\rm{d}}q \over {\rm{d}}s}=\epsilon^2 yB \;\;,\quad{{\rm{d}}r \over {\rm{d}}s}=\left(\epsilon+1\right)^2 zB   \label{dalembertfast}
\end{eqnarray}
where  $A=x^2+\epsilon^2 y^2+\left( \epsilon+1\right)^2 z^2$ and $B= p^2+q^2+r^2$. 

From these Equations, we note that $\phi={\rm{constant}}=\phi_0$ and $\omega={\rm{constant}}=\omega_0$, {\it{i.e.}} constant frequency. In addition, $t=\omega s+t_0$, where we arbitrarily set $t_0=0$, which correponds to the leading edge of the wave pulse starting at $t=0$ when $s=0$.  We can also construct the integral:
\begin{eqnarray}
{{\rm{d}} \over {\rm{d}}s} \left( xp+yq+zr \right) =0 \;\; \Rightarrow \;\; xp+yq+zr = {\rm{constant}} = x_0p_0+y_0q_0+z_0r_0\label{bbbbbb}
\end{eqnarray}
However, we are unable to find a second conserved quantity.


\subsection{Planar fast wave starting at $z_0=1$}\label{sub1}

\begin{center}
\begin{figure}[t]
\includegraphics[width=3.8in]{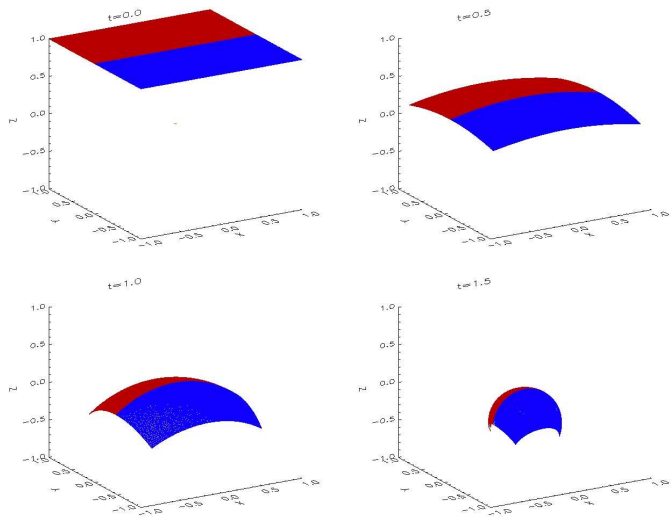}
\caption{($\epsilon=1$) Surfaces of constant $\phi$ at four values of $t$, showing the behaviour of the (initially planar) wavefront that starts at $-1 \leq x \leq 1$, $-1 \leq y <0$ and $z=1$ (blue) and  $-1 \leq x \leq 1$, $0 \leq  y \leq 1$ and $z=1$ (red).  The (arbitrary) colouring has been added to aid the reader in tracking the wave behaviour.  This figure is also available as an mpg animation.}
\label{figure2}
\end{figure}
\end{center}

We now solve Equation (\ref{dalembertfast}) subject to the initial conditions:
\begin{eqnarray}
\phi_0&=&0 \;\;,\quad \omega_0=2\pi \;\;,\quad -1\leq x_0 \leq 1 \;\;,\quad  -1 \leq y_0 \leq 1 \;\;,\quad z_0 =1  \;\;,\nonumber \\
p_0&=&0 \;\;,\quad q_0 = 0  \;\;,\quad r_0 = \omega_0 / \sqrt{x_0^2+\epsilon^2y_0^2+ \left( \epsilon +1\right)^2 z_0^2} \;\;,\label{chad}
\end{eqnarray}
where we have arbitrarily chosen $\omega_0 =2\pi$ and $\phi_0=0$. These initial conditions correspond to a planar fast wave being sent towards the null point from our upper boundary (along $z=z_0$).

Let us initially consider $\epsilon=1$ (corresponding to Figure \ref{figure1}: Left). In Figure \ref{figure2}, we have plotted surfaces of constant $\phi$, which can be thought of as defining the position of the wavefront,  at various times. Since $t=\omega s$, these correspond to different values of the parameter $s$, which quantifies distance travelled  along the characteristic curve. We can clearly see that the fast wave experiences a refraction effect towards the null point, {\it{i.e.}} propagation towards regions of lower Alfv\'en speed. A similar  refraction effect was also seen in the 2D case (\opencite{MH2004}; \citeyear{MH2006}). Thus, the  fast wave is deformed from its initial planar profile. The wave, and hence all the wave energy, eventually accumulate at the null point.

\begin{center}
\begin{figure}[t]
\includegraphics[width=4.8in]{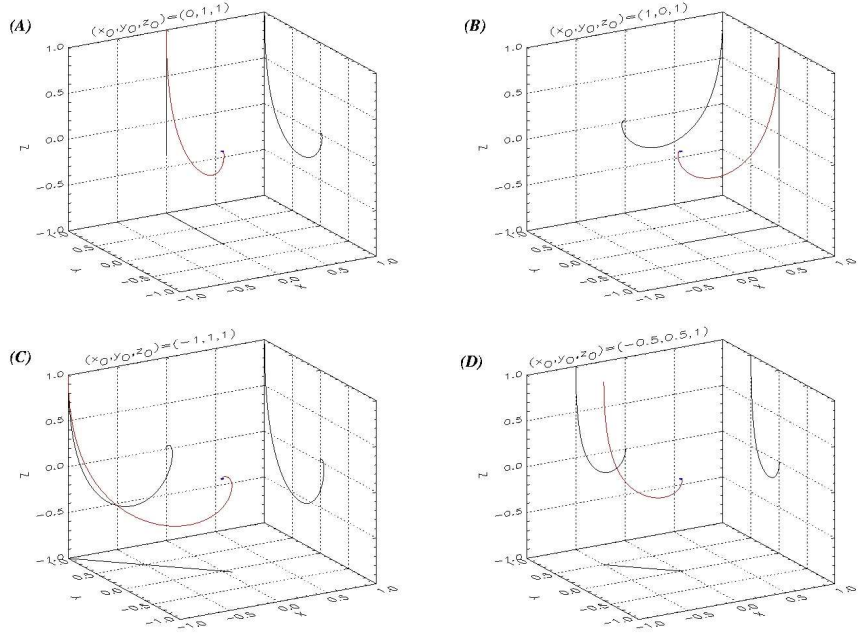}
\caption{($\epsilon=1$) Ray paths for fluid elements that begin at $(x_0,y_0,z_0)=$ $(A)$ $(0,1,1)$, $(B)$  $(1,0,1)$, $(C)$ $(-1,1,1)$ and $(D)$ $(-0.5,0.5,1)$. This figure is also available as an mpg animation  showing all $-1 \leq x_0\leq 1$, $z_0=1$ along $y=-1$ and $y=1$, and all $-1 \leq y_0\leq 1$, $z_0=1$ along $x=-1$ and $x=1$. Here, red indicates the 3D ray path and black indicates the $xy$, $yz$ and $xz$ projections of this ray path onto the respective planes.  The blue dot indicates the position of the magnetic null point.}
\label{figure3}
\end{figure}
\end{center}

\begin{center}
\begin{figure}[t]
\includegraphics[width=4.8in]{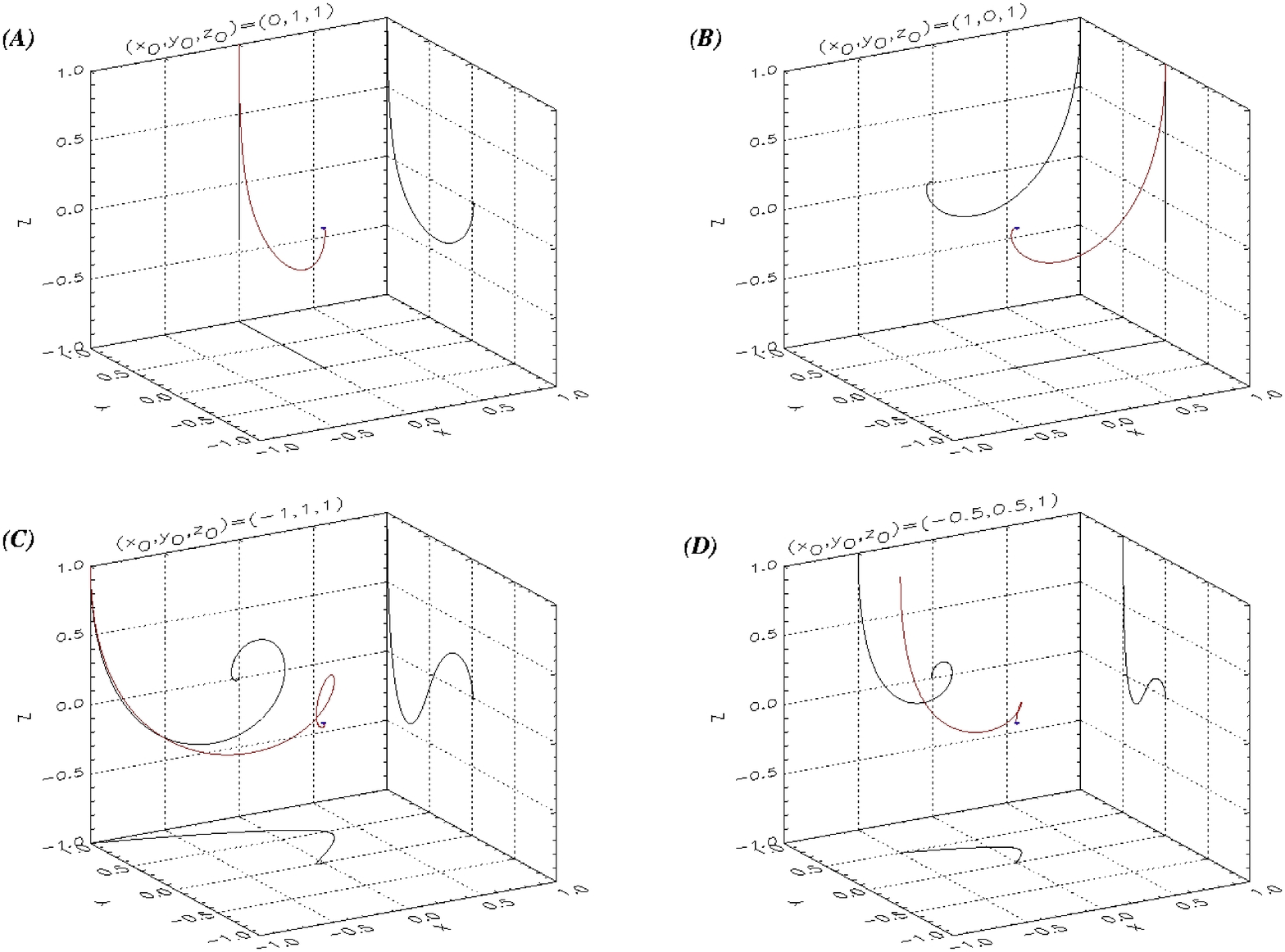}
\caption{($\epsilon=1/2$) Ray paths for fluid elements that begin at $(x_0,y_0,z_0)=$ $(A)$ $(0,1,1)$, $(B)$  $(1,0,1)$, $(C)$ $(-1,1,1)$ and $(D)$ $(-0.5,0.5,1)$. This figure is also available as an mpg animation  showing many more starting points. Red indicates the 3D ray path and black indicates the $xy$, $yz$ and $xz$ projections of this ray path onto the respective planes.  The blue dot indicates the position of the magnetic null point.}
\label{figureX}
\end{figure}
\end{center}

We can also use our WKB solution to plot the ray paths of individual fluid elements from the initial wave. In Figure \ref{figure3}, we can see the ray  paths for fluid elements that begin at four different starting points in the $z_0=1$ plane. These four ray paths are typical of the behaviour of the fast wave fluid elements (more examples can be seen in the associated mpg movie). We can clearly see the refraction effect wrapping the fast wave fluid elements around the null point.

Note that the magnitude of the refraction effect is the same for each fluid element that starts at the same radius from $z=z_0$. Thus, the $xy-$plane projections are always straight lines.  This is because the Alfv\'en speed is the same for elements starting at the same radius from $z=z_0$, {\it{i.e.}} $v_A(x_0,y_0,z_0) = \sqrt{x_0^2+y_0^2+4z_0^2}$, and the behaviour of the fast wave is entirely dominated by the Alfv\'en speed profile. For $\epsilon=1$, isosurfaces of Alfv\'en speed form prolate spheroids (parallel to the spine).

 Let us now extend our study to improper null points. In Figure \ref{figureX}, we can see the ray  paths for fluid elements that begin at the same four starting points as in Figure \ref{figure3}, but now for the magnetic field configuration seen in  Figure \ref{figure1}: Right, i.e  $\epsilon=1/2$. The first thing to note is that the refraction effect still occurs in this configuration, as expected, and that the fluid elements still eventually accumulate at the null point. However, the individual ray paths  are different to those for $\epsilon=1$.   Comparing panels \ref{figureX}$A$ and \ref{figureX}$B$ with  \ref{figure3}$A$ and \ref{figure3}$B$, we see that along $x=0$ or $y=0$, the ray paths are very similar. However, the distance travelled by the fluid element is actually different in the $\epsilon=1/2$ simulation since the Alfv\'en speeds, and hence the magnitude of the refraction, has changed.  In fact, the refraction is weaker along $y=0$ (since for $\left|B\right|=x^2+\epsilon^2y^2+ \left(\epsilon+1\right)^2z^2$, $\left|B\right|_{\epsilon=1} > \left|B\right|_{\epsilon=1/2}$ along $y=0$) and so the fluid element travels a longer distance than the equivalent $\epsilon=1$ fluid element. Along $x=0$, the effect is more complicated, with  $\left|B\right|_{\epsilon=1} > \left|B\right|_{\epsilon=1/2}$ only true for $\left|{z}\right|>\left|{y}\right|$.

The differences in the ray paths are much more obvious when comparing panels \ref{figureX}$C$ and \ref{figureX}$D$ with  \ref{figure3}$C$ and \ref{figure3}$D$. For $\epsilon=1/2$, we see that the ray paths are now \lq\lq{corkscrew}\rq{}\rq{} spirals. This is because the Alfv\'en speed profile is now varying in three directions, whereas for $\epsilon=1$ the Alfv\'en speed essentially varies in two directions: $r=\sqrt{x^2+y^2}$ and $z$. Thus, the  $xy-$plane projections are no longer straight lines. For $\epsilon=1/2$, isosurfaces of Alfv\'en speed form scalene ellipsoids.

\subsection{Planar fast wave starting at $y_0=1$}\label{sub2}

\begin{center}
\begin{figure}[t]
\includegraphics[width=3.8in]{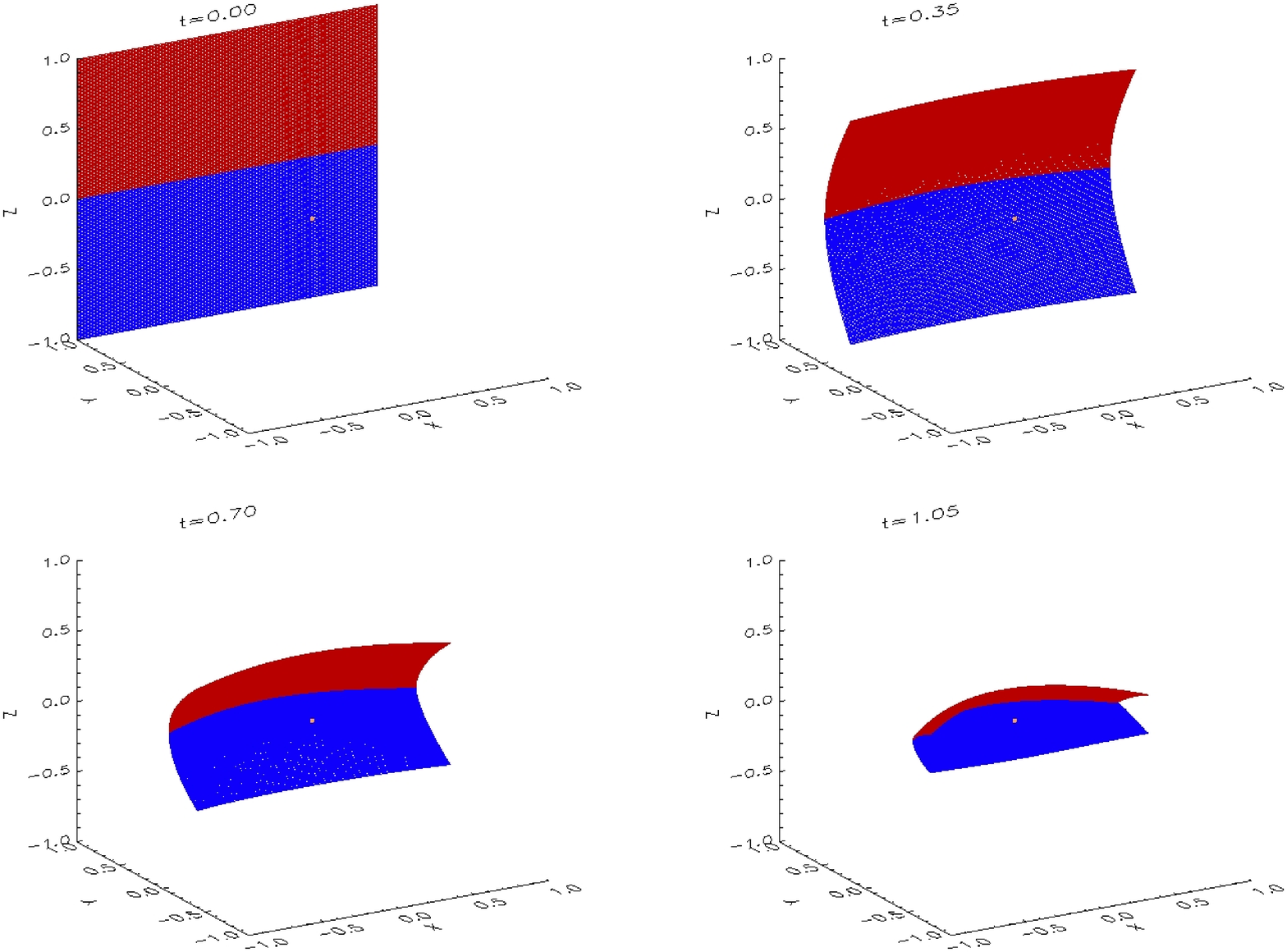}
\caption{($\epsilon=1$) Surfaces of constant $\phi$ at four values of $t$, showing the behaviour of the (initially planar) wavefront that starts at $-1 \leq x \leq 1$, $y=1$ and $-1 \leq z < 0$ (blue) and $-1 \leq x \leq 1$, $y=1$ and $ 0 \leq z \leq 1$ (red). The (arbitrary) colouring has been added to aid the reader in tracking the wave behaviour. The yellow dot indicates the position of the magnetic null point.}
\label{figure4}
\end{figure}
\end{center}

\begin{center}
\begin{figure}[t]
\includegraphics[width=4.8in]{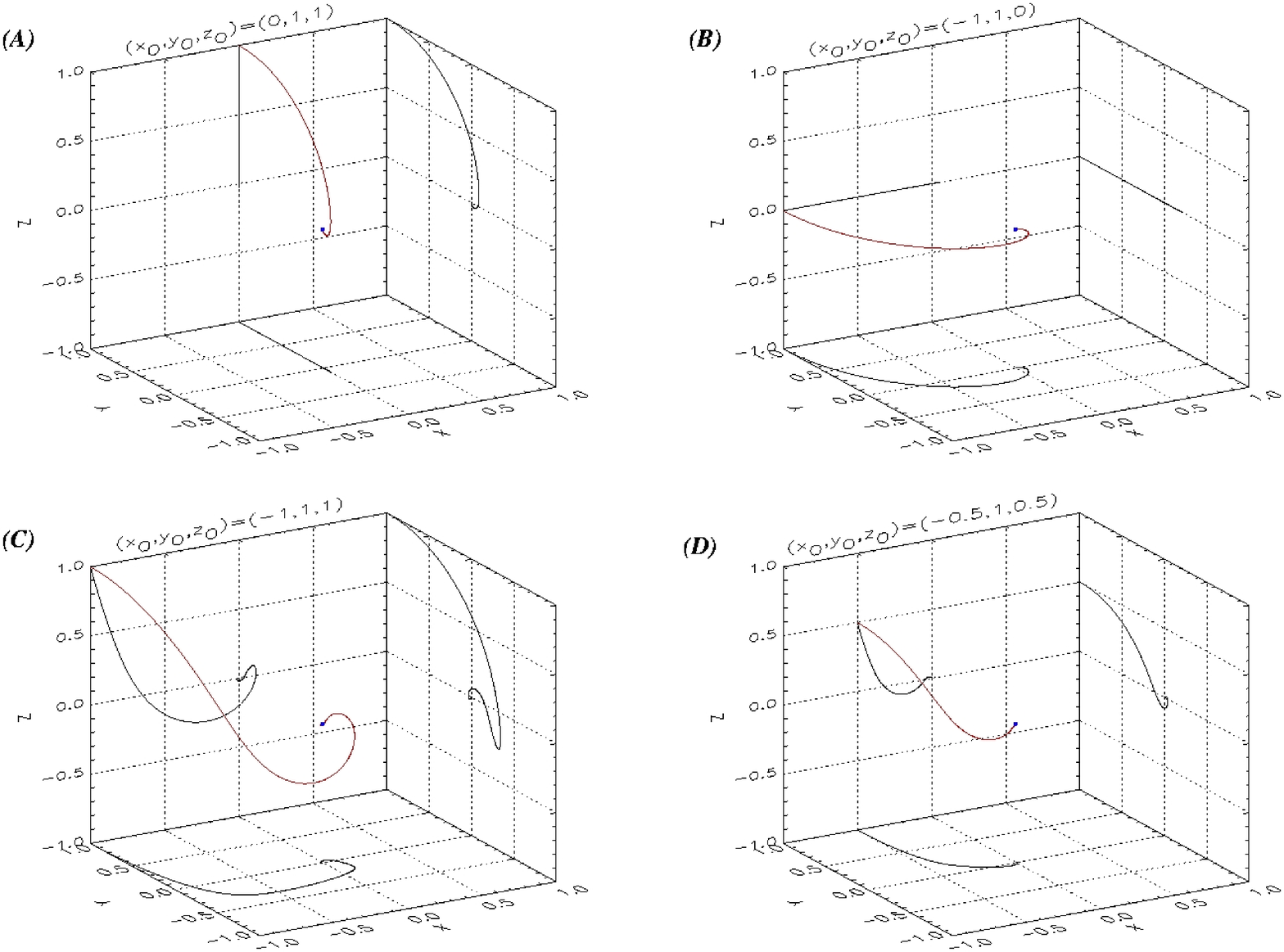}
\caption{($\epsilon=1$) Ray paths for fluid elements that begin at $(x_0,y_0,z_0)=$ $(A)$ $(0,1,1)$, $(B)$  $(-1,1,0)$, $(C)$ $(-1,1,1)$ and $(D)$ $(-0.5,1,0.5)$.  This figure is also available as an mpg animation showing many more starting points. Red indicates the 3D ray path and black indicates the $xy$, $yz$ and $xz$ projections of this ray path onto the respective planes.  The blue dot indicates the position of the magnetic null point.}
\label{figure5}
\end{figure}
\end{center}

We again solve Equation (\ref{dalembertfast}) but now subject to the initial conditions:
\begin{eqnarray}
\phi_0&=&0 \;\;,\quad \omega_0=2\pi\;\;,\quad -1 \leq x_0 \leq 1 \;\;,\quad  y_0 = 1 \;\;,\quad-1 \leq z_0 \leq 1 \;\;, \nonumber \\
p_0&=&0 \;\;,\quad q_0 = \omega_0 / \sqrt{x_0^2+\epsilon^2 y_0^2+\left( \epsilon+1\right)^2 z_0^2}  \;\;,\quad r_0 = 0 \;\;,\label{pear}
\end{eqnarray}
These initial  conditions correspond to a fast wave being sent in  from the side boundary (along $y=y_0$). This choice of planar fast wave is incident perpendicular to the spine ($z-$axis).

Let us first consider $\epsilon=1$. Figure \ref{figure4} shows surfaces of constant $\phi$ at four values of $t$,  showing the behaviour of the (initially planar) wavefront that starts at $-1 \leq x \leq 1$, $y=1$ and $-1 \leq z \leq 1$. Again, we see the deformation of the (initially planar) wave due to the refraction effect and, again, the wave accumulates at the null point. However, the nature of this refraction is different to that seen in Figure \ref{figure2}, since the refraction varies in magnitude in different planes. We see the wavefront is initially \lq\lq pinched\rq{}\rq{} preferentially in the $yz-$plane (since $ v_A(y,z)> v_A(x,y)$ for  $\left|{2z}\right|>\left|{x}\right|$).

In Figure \ref{figure5}, we can see the ray  paths for fluid elements that begins at four different starting points in the $y_0=1$ plane. Again, we can clearly see the refraction wrapping the fast wave elements around the null point, and the ray paths accumulate at the null point.

We do not show  the ray paths corresponding to a planar fast wave starting at $x_0=1$ and approaching the null point, since these the ray paths behave identically to those in Figure \ref{figure5} under the transformation $(x,y)\rightarrow  (-y,x)$ (for $\epsilon=1$ configuration). The ray paths corresponding to a planar fast wave starting at $y_0=1$ and starting at $x_0=1$ in the $\epsilon=1/2$ magnetic configuration can be found in Figures \ref{extra1} and \ref{extra2} in Appendix \ref{extra}.

Thus, Sections \ref{sub1} and \ref{sub2} have shown that the fast wave experiences a refraction effect in the neighbourhood of a 3D magnetic null point and that in all of these cases, the main result is the same: the ray paths accumulate at the null point. Of course, the actual paths taken vary depending upon initial conditions and choice of $\epsilon$. Hence, we conclude that the fast wave, and thus the fast-wave energy, eventually accumulate at the 3D null point for all $\epsilon$ and all initial conditions that generate a wave approaching the null.

Finally, it should be noted that the behaviour of the fast wave is entirely dominated by the Alfv\'en-speed profile, and since the magnetic field drops to zero at the null point,  the wave will never actually reach there. However, there is still current accumulation and hence non-ideal effects may be able to extract the wave energy in a finite time. This is investigated in the next section.


\subsection{Current build up}\label{current_fast}

From Section \ref{WKBAPPROXIMATION}, we know that for the fast wave ${\bf{v}} \cdot {\bf{B}}_0 =0$ and ${\bf{v}} \cdot \left( {\bf{B}}_0 \times {\bf{k}}\right) = 0$. This gives us two Equations for the three velocity variables:
\begin{eqnarray*}
&&\quad \quad \quad  xv_x+\epsilon y v_y - \left( \epsilon+1 \right) z v_z =0\;\; ,\nonumber \\
&&\left[ \epsilon y r + \left( \epsilon+1 \right) z q \right] v_x - \left[ xr + \left( \epsilon+1 \right) zp \right] v_y +\left[ xq - \epsilon y p \right]v_z  =0\;\; .
\end{eqnarray*}
Thus, we can express two of the velocity components in terms of the third.

Recall from Section \ref{SEC:1} that the perturbed electric current is given by ${\bf{j}}_1 = \nabla \times {\bf{B}}_1$.  Thus,
\begin{eqnarray*}
{\partial\over \partial t}  {{\bf B}}_1  &=& \nabla \times \left ({\bf v} \times {\bf B}_{0} \right) \;\; \Rightarrow \;\;  -\omega {{\bf B}}_1 =  {{\bf k}} \times \left( {{\bf v}} \times {{\bf B}}_0 \right)    \;\;, \\
\Rightarrow  {\bf{j}}_1 &=& {\rm{i}}  {{\bf k}} \times  {{\bf B}}_1 = -{\rm{i}}  {{\bf k}} \times \left[  {{\bf k}} \times \left( {{\bf v}} \times {{\bf B}}_0 \right) \right]  / \omega  = {\rm{i}} \left| {{\bf k}} \right|^2 \left( {{\bf v}} \times {{\bf B}}_0 \right)/ \omega \;\;,
\end{eqnarray*}
where we have made use of ${\bf{v}} \cdot \left( {\bf{B}}_0 \times {\bf{k}}\right) = 0$. From Equation (\ref{fastEquation}), we can substitute for $\left| {{\bf k}} \right|^2$ to obtain:
\begin{eqnarray}
 {\bf{j}}_1 &=& {\rm{i}} \omega  \left( {{\bf v}} \times {{\bf B}}_0 \right) / {\left| {{\bf B}}_0 \right|^2} \nonumber \\
&=& {\rm{i}} \omega  \frac{  \left[ - \left(\epsilon+1 \right) z v_y - \epsilon y v_z, \left( \epsilon +1 \right) z v_x + xv_z, \epsilon y v_x - xv_y \right] }{ x^2 + \epsilon ^2 y^2 + \left( \epsilon +1 \right)^2 z^2} \;\;,\nonumber\\
 \Rightarrow  \left|{\bf{j}}_1 \right|&=&  \omega  \left|{\bf{v}} \right| /   \left|{\bf{B}}_0 \right| \;\;, \label{ALANcurrent} 
\end{eqnarray}
where we have used ${\bf{v}} \cdot {\bf{B}}_0 =0$ to simplify $\left| {{\bf v}} \times {{\bf B}}_0 \right|$. Thus, since $\left| {{\bf v}} \right|$ is bounded (from our assumed form of  ${{\bf v}}$ seen in Equation \ref{WKB}) we can see that the current associated with the fast wave will grow $\sim 1/ \left| {{\bf B}}_0 \right|$. Equivalent behaviour was found for the fast wave in the 2D case \cite{MH2004}, {\it{i.e.}} $\epsilon=0$.

Moreover, we can place limits on the magnitude of the current build up. From Equation (\ref{kaj}) in Appendix \ref{limits}, we can place limits on $\left|{\bf{j}}_1 \right|$ such that:
\begin{eqnarray}
\frac{  \omega    \left|{\bf{v}} \right|   }{  \left(\epsilon+1\right)  R_0  } {\rm{e}}^{  \alpha \epsilon^2 t / \omega_0}  \le \left|{\bf{j}}_1\right|  \le  \frac{  \omega   \left|{\bf{v}} \right|        }{\epsilon  \:    R_0 }    {\rm{e}}^{  \alpha \left(\epsilon+1\right)^2  t / \omega_0} \;\;,\label{dave}  
\end{eqnarray}
where $R_0^2=x_0^2+y_0^2+z_0^2$ and $\alpha=x_0p_0+y_0q_0+z_0r_0$ (see Appendix \ref{limits}) and we have assumed $0\le \epsilon \le1$. Thus, we can see that the current build up is bounded by two exponentially growing functions.

We now demonstrate this current build up for two particular cases. Firstly, consider a planar fast wave starting at $z_0=1$ (Section \ref{sub1}). Here, we can solve Charpit's Equations for the fast wave (Equation \ref{dalembertfast}) analytically for the initial conditions  $p_0=q_0=x_0=y_0=0$, {\it{i.e.}} along $x=y=0$ which is the path along which we expect the maximum current build up to occur. Under these conditions, Equation  (\ref{dalembertfast}) reduces to:
\begin{eqnarray*}
x=0\;\;,\;\; y=0\;\;,\quad{{\rm{d}}z \over {\rm{d}}s}=- \left(\epsilon+1\right)^2  r z^2\;\;,\;\; p=0\;\;,\;\; q=0\;\;, \quad {{\rm{d}}r \over {\rm{d}}s}=\left(\epsilon+1\right)^2 z r^2  \;\;,
\end{eqnarray*}
where we have used initial conditions (\ref{chad}). We also note that our conserved quantity (Equation \ref{bbbbbb}) states $zr=z_0r_0=\omega_0 / \left(\epsilon+1\right)$. Thus:
\begin{eqnarray}
z=z_0 {\rm{e}}^{-\left(\epsilon+1\right) \omega_0 s} =z_0 {\rm{e}}^{-\left(\epsilon+1\right) t}    \;\;,\quad        r = r_0 {\rm{e}}^{\left(\epsilon+1\right) \omega_0 s}=  \frac{\omega_0}{\left(\epsilon+1\right)z_0}  {\rm{e}}^{\left(\epsilon+1\right) t}\label{chris}\;\;.
\end{eqnarray}

As mentioned previously, the Alfv\'en speed drops to zero at the null point, indicating that the wave will never actually reach there, but the length scales (this can be thought of as the distance between the leading and trailing edges of the wave pulse) rapidly decrease, indicating that the current (and all other gradients) will increase.  As an illustration, consider the wavefront as it propagates down the $z-$axis along $x=y=0$. From Equation (\ref{chris}), the leading edge of the wave pulse is located at a position $z=z_0 {\rm{e}}^{-\left(\epsilon+1\right) t}$, when the wave is initally at $z=z_0$. If the trailing edge of the wave pulse leaves $z=z_0$ at $t=t_1$ then the location of the trailing edge of the wave pulse at a later time is  $z_2=z_0 {{\rm{e}}}^{-\left(\epsilon+1\right)  \left(t-t_1\right)}$. Thus, the distance between the leading and trailing edges of the wave is $\delta z = z_0 {\rm{e}}^{-\left(\epsilon+1\right)   t} \left( {\rm{e}}^{\left(\epsilon+1\right)  t_1} -1 \right)$ and this decreases with time, suggesting that all gradients will increase exponentially.

We can also find analytical solutions for the velocity and polarisation of the fast wave.  ${\bf{v}} \cdot {\bf{B}}_0 =0$ along $x=y=0$ implies $v_z=0$, and hence using Equation (\ref{chris}) we obtain:
\begin{eqnarray*}
{\bf{k}} = \left(0,0,r_0 {\rm{e}}^{\left( \epsilon +1 \right) t }\right) \;\;,\quad {\bf{v}} = \left( v_x,v_y,0  \right)  {\rm{e}}^{{\rm{i}} \phi_0 }\;\;. \label{apple}
\end{eqnarray*}
Using these forms in Equation (\ref{ALANcurrent}) gives: 
\begin{eqnarray}
 {\bf{j}}_1  = - \frac{{\rm{i}} \omega_0} {\left( \epsilon +1 \right) z} \left( -  v_y ,  v_x,0 \right)   {\rm{e}}^{{\rm{i}} \phi_0 }  = -\frac{ {\rm{i}}  \omega_0}{\left( \epsilon +1 \right) z_0}  {\rm{e}}^{\left( \epsilon +1 \right) t}   \left( v_y, -v_x,0\right)  {\rm{e}}^{{\rm{i}} \phi_0 }  \;\;,\label{example}
\end{eqnarray}
where we have substituted for $z$  from Equation (\ref{chris}). Thus, along the $z-$axis current builds up exponentially: $\left|{\bf{j}}_1\right|\sim z^{-1} \sim  {\rm{e}}^{\left( \epsilon +1 \right) t}$. Comparing to Equation (\ref{dave}) we see that this exponent is the same as that of our theoretical maximum current build up (under these  initial conditions  $\alpha =\omega_0 / \left(\epsilon+1\right)$). The coefficient is slightly smaller than our theoretical maximum, but this is most likely because the limits we assumed for Equation (\ref{ant}) (see Appendix \ref{limits}) were not very strong.

Secondly, for a planar fast wave starting at $y=y_0$ (Section \ref{sub2}), we can perform the same analysis along $x=z=0$.  Using the appropriate initial conditions (Equation \ref{pear}) and following the same analysis as above, we obtain:
\begin{eqnarray*}
y&=&y_0  {\rm{e}}^{-\epsilon t}\;\;,\quad {\bf{v}}=\left( v_x,0,v_z\right)  {\rm{e}}^{{\rm{i}} \phi_0 } \;\;,\quad {\bf{k}}=\left( 0,q_0  {\rm{e}}^{\epsilon t},0\right)\\
\Rightarrow {\bf{j}}_1 &=&  -\frac { {\rm{i}} \omega_0   }{\epsilon y_0}  {\rm{e}}^{\epsilon t}   \left( v_z, 0, -v_x\right)   {\rm{e}}^{{\rm{i}} \phi_0 } \;\;.
\end{eqnarray*}
Hence, we have exponential current build up: $\left|{\bf{j}}_1\right|\sim y^{-1} \sim  {\rm{e}}^{\epsilon t}$.  Again, this exponential build up is within our theoretical limits (Equation \ref{dave}).


\section{Alfv\'en Wave}\label{SEC:ALFVEN}

\begin{center}
\begin{figure}[t]
\includegraphics[width=3.8in]{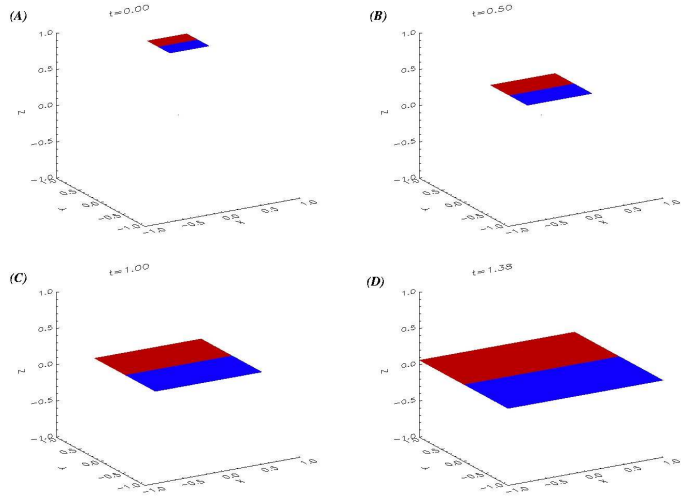}
\caption{($\epsilon=1$) Surfaces of constant $\phi$ at four values of $t$, showing the behaviour of the (initially planar) wavefront that starts at $-0.25 \leq x \leq 0.25$, $-0.25 \leq y < 0$, $z_0=1$  (blue) and $-0.25 \leq x \leq 0.25$, $0 \leq y \leq 0.25$, $z_0=1$  (red).  The (arbitrary) colouring has been added to aid the reader in tracking the wave behaviour.}
\label{figure6}
\end{figure}
\end{center}


We now consider the second root to Equation (\ref{F1}) which corresponds to the Alfv\'en wave. Hence, we assume  $\omega^{2} \neq  \left| {\bf{B}}_0   \right| ^2  \left|   {\bf{k}} \right| ^2$ and simplify Equation (\ref{F1}) to:
\begin{eqnarray}
{\mathcal{F}} \left( \phi, x,y,z,p,q,r \right) &=&  \omega^2 - \left( {\bf{B}}_0 \cdot {\bf{k}} \right) ^{2} \nonumber\\
  &=& \omega^2 - \left( xp+\epsilon yq-\left(\epsilon +1\right) zr \right)^2=0      \;\;.     \label{alfvenEquation}
\end{eqnarray}

Charpit's Equations relevant to Equation (\ref{alfvenEquation}) are:
\begin{eqnarray}
{{\rm{d}}\phi \over {\rm{d}}s}&=&0 \;\;,\quad{{\rm{d}} t \over {\rm{d}}s}=\omega \;\;,\quad{{\rm{d}}x \over {\rm{d}}s}=-x\xi \;\;,\quad{{\rm{d}}y \over {\rm{d}}s}=- \epsilon y\xi \;\;,\quad{{\rm{d}}z \over {\rm{d}}s}= \left( \epsilon +1 \right) z \xi\;\;, \nonumber\\
{{\rm{d}}\omega \over {\rm{d}}s}&=&0 \;\;,\quad{{\rm{d}}p \over {\rm{d}}s}=\;p\xi \;\;,\quad{{\rm{d}}q \over {\rm{d}}s}=\epsilon q \xi \;\;,\quad{{\rm{d}}r \over {\rm{d}}s}=- \left( \epsilon +1 \right) r\xi\;\;,\label{dalembert}
\end{eqnarray}
where $\xi = xp+\epsilon yq- \left( \epsilon +1 \right) zr$. Thus, we can see that $\phi={\rm{constant}}=\phi_0$ and $\omega={\rm{constant}}=\omega_0$. In addition, $t=\omega s$, where we have set $t=0$ at $s=0$.

\subsection{Planar Alfv\'en Wave starting at $z_0=1$}\label{sub3}

\begin{center}
\begin{figure}[t]
\includegraphics[width=4.8in]{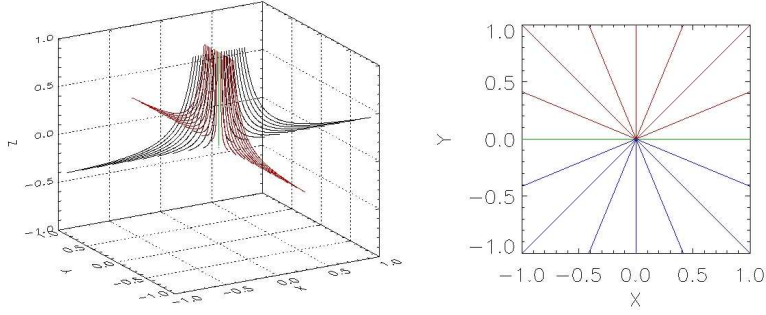}
\caption{($\epsilon=1$) {\emph{Left}}: Ray paths for fluid elements that begin at points $-0.25 \leq x_0\leq 0.25$ along $y_0=0$, $z_0=1$ (indicated in black), and $-0.25 \leq y \leq 0.25$ along $x_0=0$, $z_0=1$ (indicated in red) after a  time $t=\pi/2$. The ray path from $x_0=y_0=0$ is indicated in green and corresponds to the spine fieldline. {\emph{Right}}: Projection of ray paths onto the $xy-$plane (red indicates $y>0$, blue $y<0$ and green $y=0$.}
\label{figure7}
\end{figure}
\end{center}

We now solve Equation (\ref{dalembert}) as before, subject to the initial conditions:
\begin{eqnarray}
\phi_0&=&0 \;\;,\quad \omega_0=2\pi \;\;,\quad -1 \leq x_0 \leq 1 \;\;,\quad  -1 \leq y_0 \leq 1 \;\;,\quad z_0 =1 \;\;, \nonumber\\
p_0&=&0 \;\;,\quad q_0 = 0  \;\;,\quad r_0 = \omega_0 / { \left[ \left( \epsilon +1 \right) z_0  \right]}\label{KKK}  \;\;,
\end{eqnarray}
where we have (arbitrarily) chosen $\omega_0=2\pi$ and $\phi_0=0$. This corresponds to a planar Alfv\'en wave initially at $z=z_0$. 


We can see the behaviour of the Alfv\'en wavefront in Figure \ref{figure6} (we have plotted surfaces of constant $\phi$ as in Section \ref{sub1}). We have also only plotted the wavefronts originating from  $-0.25 \leq x_0,y_0 \leq 0.25$ so as to better illustrate the wavefront evolution. We can clearly see that the initially planar wavefront expands (in the $xy-$plane) as it approaches the null point, and keeps its original shape ({\it{i.e.}} planar and no rotation). The Alfv\'en wave eventually accumulates along the fan plane, and never enters the $z<0$ domain.


In Figure \ref{figure7}: Left, we can see the ray paths for fluid elements that begin at points $-0.25 \leq x_0\leq 0.25$, $y_0=0$, $z_0=1$  and $-0.25 \leq y_0 \leq 0.25$,  $x_0=0$, $z_0=2$,  after a  time $t=\pi/2$. Here, we see that the fluid elements travel along and are confined to the fieldlines they start on, {\it{i.e.}} the Alfv\'en wave spreads out following the fieldlines. This explains the expansion of the wavefront seen in Figure \ref{figure6}. A similar effect was seen in the 2D case \cite{MH2004}. As noted for the wavefront, all the elements have travelled a different distance along their respective fieldlines but still form a planar wave. This is explained in section \ref{appendixB}.

\subsection{Planar Alfv\'en Wave starting at $y_0=1$}\label{sub4}

\begin{center}
\begin{figure}[t]
\includegraphics[width=3.8in]{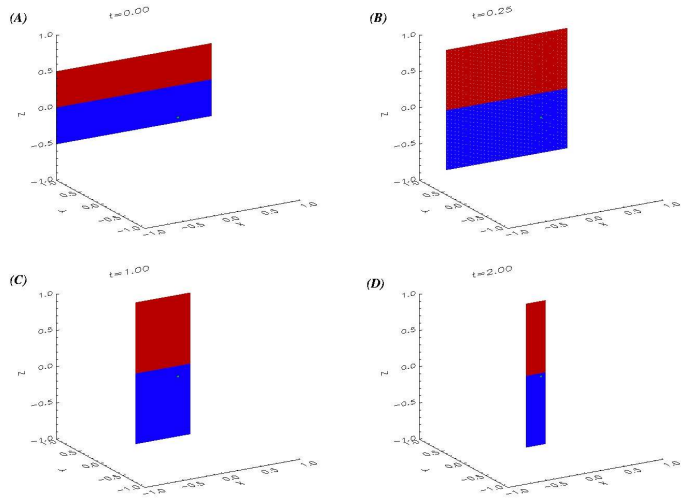}
\caption{($\epsilon=1$) Surfaces of constant $\phi$ at four values of $t$, showing the behaviour of the (initially planar) wavefront that starts at $-1\leq x_0 \leq 1$, $y_0=1$ and $-0.5 \leq z_0 <0$ (blue) and $-1\leq x_0 \leq 1$, $y_0=1$ and $0 \leq z_0 \leq 0.5$ (red). The (arbitrary) colouring has been added to aid the reader in tracking the wave behaviour. The green dot indicates the position of the magnetic null point.  We have imposed maximum and minimum values of unity in the $z-$direction, purely for illustrative purposes.  }
\label{figure8}
\end{figure}
\end{center}

We again solve Equation (\ref{alfvenEquation}) but now subject to the initial conditions:
\begin{eqnarray}
\phi_0&=&0 \;\;,\quad \omega_0=2\pi \;\;,\quad -1 \leq x_0 \leq 1 \;\;,\quad  y_0 = 1 \;\;,\quad-1 \leq z_0 \leq 1\;\;,\nonumber\\
p_0&=&0 \;\;,\quad q_0 =\omega_0 / \left( \epsilon y_0\right)   \;\;,\quad r_0 =0 \;\;.\label{222}
\end{eqnarray}
 This corresponds to an Alfv\'en wave being sent in from the side boundary (along $y=y_0$).

We can see the behaviour of the Alfv\'en wavefront in Figure \ref{figure8} (surfaces of constant $\phi$).  We have plotted the wavefronts starting at $-1\leq x_0 \leq 1$, $y_0=1$ and $-0.5 \leq z_0 \leq 0.5$ in order to more clearly show the Alfv\'en wave propagation. We can see that the (initially rectangular) wavefront expands in the $z-$direction but is also squeezed in the $x-$direction as it approaches the null ({\it{i.e.}} as $y$ decreases). We have imposed maximum and minimum values of unity in the $z-$direction, purely for illustrative purposes.  The Alfv\'en wave, and hence the wave energy, eventually accumulates along the spine. Again, the wave remains planar as it propagates.

\begin{center}
\begin{figure}[t]
\includegraphics[width=4.8in]{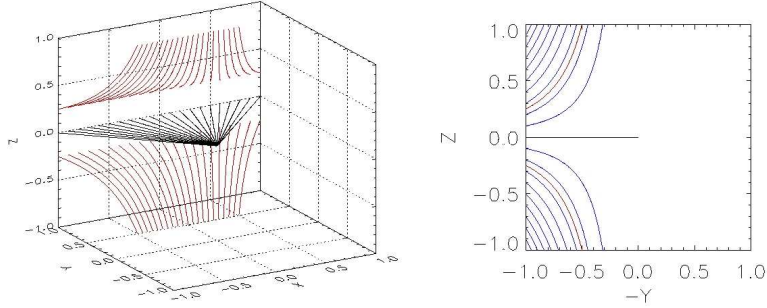}
\caption{($\epsilon=1$) {\emph{Left}}: Ray paths for fluid elements that begin at points $-1 \leq x_0 \leq 1$, $y_0=1$ and $z_0=\pm0.25$ (indicated in red) and $z_0=0$ indicated in black) after time $t=2\pi$.  We have imposed maximum and minimum values of unity in the $z-$direction, purely for illustrative purposes.   {\emph{Right}}: Ray paths  for fluid elements that begin at points $-1 \leq z_0 \leq 1$  in  the $yz-$plane along $x=0$ (blue indicates starting points of $-1\le z_0\le 1$ in divisions of $0.1$, red indicates   $z_0=\pm0.25$, black indicates $z_0=0$). We have also plotted $y\rightarrow -y$ to aid the comparison between the left and right figures.}
\label{figure9}
\end{figure}
\end{center}

In Figure \ref{figure9}: Left, we can see the ray  paths for fluid elements that begin at points $-1 \leq x_0 \leq 1$, $y_0=1$ and at $z=-0.25, 0, 0.25$ after time $t=2\pi$, where we have imposed maximum and minimum values of unity in the $z-$direction (again purely for illustrative purposes). The ray paths in the fan plane all focus towards the null, which is expected as they follow the fan-fieldlines. In contrast, the fluid elements on fieldlines above and below the fan plane propagate away from the null point, but are simply following their respectively fieldlines. This is also clearly seen in Figure \ref{figure9}: Right, which shows various ray paths in the $yz-$plane along $x=0$. This behaviour explains the narrowing and stretching effect seen in Figure \ref{figure8}: the Alfv\'en wave  crosses the fan plane in this scenario and thus travels along the radially converging fan plane fieldlines. Meanwhile, the stretching effect comes from  the diverging fieldlines the wave initially crosses.  This work highlights the importance of understanding the magnetic topology of  a system.


\section{Analytical Solution for Alfv\'en wave}\label{appendixB}

We can also solve Charpit's Equations for the Alfv\'en wave (\ref{dalembert}) analytically. Firstly, let us consider a  planar wave starting at $z=z_0$. Using the appropriate  initial conditions (Equation \ref{KKK}), we find:
\begin{eqnarray}
{{\rm{d}}\xi \over {\rm{d}}s}&=& {{\rm{d}} \over {\rm{d}}s}\left( xp+\epsilon yq- \left( \epsilon +1 \right) zr\right) =0\nonumber\\
\Rightarrow \quad\xi &=& x_0p_0+ \epsilon  y_0q_0- \left( \epsilon +1 \right) z_0r_0 = -\omega_0 \label{peter}
\end{eqnarray}
where $\xi = xp+\epsilon yq- \left( \epsilon +1 \right) zr$ as before, and where the values of  $x_0$, $p_0$, $y_0$, $q_0$, $z_0$, $r_0$ and the sign of $\omega_0$ are taken from Equation (\ref{KKK}). Thus, Equation (\ref{dalembert}) can be solved analytically:
\begin{eqnarray}
p&=&p_0{\rm{e}}^{-t} \;\;,\quad q=q_0{\rm{e}}^{-\epsilon t} \;\;,\quad r=r_0{\rm{e}}^{ \left( \epsilon +1 \right) t} \;\;\;,\nonumber\\
x&=&x_0{\rm{e}}^{t} \;\;\;\;,\quad y=y_0{\rm{e}}^{ \epsilon  t} \;\;\;\;,\quad z=z_0{\rm{e}}^{- \left( \epsilon +1 \right) t}\;\;.\label{hhhh}
\end{eqnarray}
where $t=\omega_0 s$. This solution is valid for all $\epsilon$. For $\epsilon=0$, we recover the 2D solution of \inlinecite{MH2004}.

From these Equations we can see why an initially planar wave remains planar: if $p$  and $q$ are initially zero ($p_0=q_0=0$) they remain zero for all time. In addition, $z$ is independent of starting position  $x_0$ and $y_0$. Thus, after a given time, different elements have travelled different distances along their respective fieldlines, but all have the same $z$, {\it{i.e.}} all remain planar if originally planar. In addition,  it can be shown that the volume occupied by the Alfv\'en wave pulse is conserved (Appendix \ref{Volume}).

Consider a circular wavefront at $z=z_0$, such that $x_0^2+y_0^2=r^2$, where $r$ is some chosen radius. Let $x_1$, $y_1$, $z_1$ represent the position of the wavefront after some time $t$. Thus, the change in length scales ($\delta {\bf{x}}$) can be represented as:
\begin{eqnarray*}
\delta x = \left( x_0-x_1\right) {\rm{e}}^{t}\;\;,\quad \delta y = \left( y_0-y_1\right) {\rm{e}}^{\epsilon t}\;\;,\quad \delta z = \left( z_0-z_1\right) {\rm{e}}^{-\left(\epsilon +1 \right) t}\;\;.
\end{eqnarray*}
Thus, the wave eventually accumulates along the fan plane, {\it{i.e.}} $\delta x\rightarrow \infty$, $\delta y\rightarrow\infty$, $\delta  z\rightarrow 0$. Furthermore, the circular wavefront evolves as:
\begin{eqnarray*}
x_0^2+y_0^2=r^2 \quad \Rightarrow \quad\left( \frac {x}{{\rm{e}}^{t}} \right)^2 +  \left( \frac {y}{{\rm{e}}^{\epsilon t} }\right)^2 = r^2\;\;,
\end{eqnarray*}
{\it{i.e.}} the circle becomes an ellipse (with semimajor-axis in the direction of $x$ if $0<\epsilon<1$, $y$ if $\epsilon>1$ and remains circular for $\epsilon=1$). Thus, the wave only accumulates over the whole fan plane for $\epsilon=1$ and instead accumulates along a preferential axis for $\epsilon \neq 1$.

Charpit's Equations (Equation \ref{dalembert}) can also be solved using the initial conditions for a planar wave starting at $y=y_0$, {\it{i.e.}} Equation (\ref{222}). Following the same techniques above,  we see that the length scales evolve as:
\begin{eqnarray*}
\delta x = \left( x_0-x_1\right) {\rm{e}}^{-t}\;\;,\quad \delta y = \left( y_0-y_1\right) {\rm{e}}^{-\epsilon t}\;\;,\quad \delta z = \left( z_0-z_1\right) {\rm{e}}^{\left(\epsilon +1 \right) t}\;\;.
\end{eqnarray*}
In this case, the wave eventually accumulates along the spine, {\it{i.e.}} $\delta x\rightarrow 0$, $\delta y\rightarrow 0$, $\delta z\rightarrow\infty$, for all values of $\epsilon$. As before, an initially circular wavefront becomes elliptical for $0< \epsilon \neq 1$, and evolves according to:
\begin{eqnarray*}
x_0^2+z_0^2=r^2 \quad \Rightarrow \left( \frac {x}{{\rm{e}}^{-t} }\right)^2 +  \left( \frac {z}{{\rm{e}}^{\left(\epsilon+1\right)t} }\right)^2 = r^2\;\;.
\end{eqnarray*}

\subsection{Wavevector and Velocity}\label{Polarisation}

From Section \ref{WKBAPPROXIMATION}, we know that for the Alfv\'en wave ${\bf{v}} \cdot {\bf{B}}_0 =0$, ${\bf{v}} \cdot {\bf{k}} =0$ and ${\bf{v}} \cdot \left( {\bf{B}}_0 \times {\bf{k}}\right) \neq 0$. Consider a planar wave starting at $z=z_0$, using Equation (\ref{hhhh}) we obtain:
\begin{eqnarray*}
v_x x_0{\rm{e}}^{t} + v_y \epsilon  y_0{\rm{e}}^{ \epsilon  t} - v_z \left( \epsilon +1 \right)  z_0{\rm{e}}^{- \left( \epsilon +1 \right) t}&=&0\\
v_x p_0{\rm{e}}^{-t} + v_y q_0 {\rm{e}}^{ -\epsilon  t} + v_z r_0{\rm{e}}^{ \left( \epsilon +1 \right) t} &=&0
\end{eqnarray*}
 Using the  initial conditions from Equation (\ref{KKK}), $p_0=q_0=0$ and so $v_z=0$. Thus:
\begin{eqnarray}
{\bf{k}} = \left(0,0,r_0 {\rm{e}}^{\left( \epsilon +1 \right) t }\right) \;\;,\quad {\bf{v}} = v_y \left(-\frac{y_0}{x_0} {\rm{e}}^{\left( \epsilon -1 \right) t},1,0\right)  {\rm{e}}^{{\rm{i}} \phi_0 } \;\;. \label{v1}
\end{eqnarray}
Thus, the angle between $v_x$ and $v_y$ changes with time. There is one special case: for $\epsilon=1$, we have $v_x/ v_y = -y/x = -\tan{\theta}$. Recall in cylindrical coordinates $v_x= v_r \cos{\theta} - v_\theta \sin{\theta}$,  $v_y= v_r \sin{\theta} + v_\theta \cos{\theta}$, and so we must have $v_r=0$ and $v_\theta \neq0$. Hence, for $\epsilon=1$  we have circular rotation of the fieldlines.

Similarly, for a planar Alfv\'en wave starting at $y=y_0$ (Equation \ref{222}) and using the same derivation as above, we obtain:
\begin{eqnarray*}
{\bf{k}} = \left(0,q_0 {\rm{e}}^{\epsilon t },0\right) \;\;,\quad {\bf{v}} = v_z \left( \left(\epsilon +1\right) \frac{z_0}{x_0} {\rm{e}}^{\left( \epsilon +2 \right) t},0,1\right)  {\rm{e}}^{{\rm{i}} \phi_0 }\;\;.
\end{eqnarray*}

Finally, for a planar Alfv\'en wave starting at $x=x_0$, we obtain:
\begin{eqnarray*}
{\bf{k}} = \left(p_0{\rm{e}}^{ t },0,0\right) \;\;,\quad {\bf{v}} = v_z   \left(  0,  \frac{ \epsilon+1}{\epsilon } \frac{ z_0}{y_0} {\rm{e}}^{\left( 2\epsilon +1 \right) t}, 1 \right){\rm{e}}^{{\rm{i}} \phi_0 } \;\;.
\end{eqnarray*}
These velocity and polarisation solutions will be used in the next section.

\subsection{Current build up}

Recall from Section \ref{SEC:1} that the perturbed electric current is given by ${\bf{j}}_1 = \nabla \times {\bf{B}}_1$. Now that we have an analytic solution for ${\bf{v}}$ we can solve Equation (\ref{6}) for  ${\bf{B}}_1$. Hence, ${\bf{j}}_1$ can be found:
\begin{eqnarray*}
{\partial\over \partial t}  {{\bf B}}_1  &=& \nabla \times \left ({\bf v} \times {\bf B}_{0} \right) \;\; \Rightarrow \;\;  -\omega_0 {{\bf B}}_1 =  {{\bf k}} \times \left( {{\bf v}} \times {{\bf B}}_0 \right)  =  \left( {{\bf B}}_0 \cdot {{\bf k}} \right){{\bf v}} - \left( {{\bf k}} \cdot {{\bf v}} \right) {{\bf B}}_0 \;\;, \\
&\Rightarrow&  {\bf{j}}_1 = {\rm{i}}  {{\bf k}} \times  {{\bf B}}_1 = -{\rm{i}} \left( {\bf{k}} \times {{\bf v}} \right)  \left( {{\bf B}}_0 \cdot {{\bf k}} \right) / \omega_0   \;\;,
\end{eqnarray*}
where we have made use of ${\bf {v}} \cdot {{\bf k}} =0$. Let us first consider a planar wave starting at $z=z_0$. Using the forms of ${\bf{v}}$ and ${\bf{k}}$ from Equation (\ref{v1}) gives:
\begin{eqnarray}
 {\bf{j}}_1  =  -{\rm{i}} \left( {\bf{k}} \times {{\bf v}} \right)   \xi / \omega_0 =  -  \frac{ {\rm{i}} \omega_0}{\left(\epsilon+1 \right) x_0z_0} v_y  \left(   x_0 {\rm{e}}^{\left( \epsilon +1 \right) t },  y_0  {\rm{e}}^{2 \epsilon  t},0 \right)   {\rm{e}}^{{\rm{i}} \phi_0 } \;\;,\label{example2}
\end{eqnarray}
where $\xi={{\bf B}}_0 \cdot {{\bf k}}=-\omega_0$ and $\omega_0=\left( \epsilon+1\right) z_0r_0$ from Equation (\ref{peter}).

Thus, we have an exponential build up of $j_x$ and $j_y$ in our system.  For $\epsilon=0$, this reduces to $j_x \sim   {\rm{e}}^{t}$ as found by \inlinecite{MH2004}. We also see that the current build up is the fan-plane.

Similarly, for a planar Alfv\'en wave starting at $y=y_0$, we obtain:
\begin{eqnarray}
 {\bf{j}}_1=    - \frac{ {\rm{i}}  \omega_0}{\epsilon x_0 y_0} v_z  \left(  x_0          {\rm{e}}^{ \epsilon t }, 0, -\left(\epsilon +1\right) z_0   {\rm{e}}^{2 \left(\epsilon +1\right) t} \right)   {\rm{e}}^{{\rm{i}} \phi_0 } \;\;,\label{FFF}
\end{eqnarray}
where $\xi={{\bf B}}_0 \cdot {{\bf k}}=\omega_0$ and $\omega_0=  \epsilon y_0 q_0$ (from Equation \ref{222}).

Finally, for a planar wave starting at $x=x_0$, we obtain:
\begin{eqnarray}
 {\bf{j}}_1=   -\frac{ {\rm{i}} \omega_0}{\epsilon x_0 y_0} v_z  \left(0,            -  \epsilon  y_0   {\rm{e}}^{\epsilon t}, \left(\epsilon+1 \right) z_0 {\rm{e}}^{ 2\left( \epsilon + 1 \right) t } \right) {\rm{e}}^{{\rm{i}} \phi_0 } \;\;,\label{GGG}
\end{eqnarray}
where $\xi={{\bf B}}_0 \cdot {{\bf k}}=\omega_0$ and $\omega_0= x_0 p_0$. For $\epsilon=1$, Equation (\ref{GGG}) is the same as Equation (\ref{FFF}) under the transformation $(x,y)\rightarrow (-y,x)$. We can see that for both Equations  (\ref{FFF}) and  (\ref{GGG}), the current build up is predominately along the spine.


\section{Conclusion}\label{conclusion}

We have demonstrated  how the WKB approximation can be used to help solve the linearised MHD Equations. Using Charpit's Method and a Runge-Kutta numerical scheme, we have demonstrated this technique for a general 3D potential magnetic null point (parameter $\epsilon$).  Under the assumptions of ideal and cold plasma, we have considered two types of wave propagation: fast magnetoacoustic and Alfv\'enic.

For the fast magnetoacoustic wave, we find that the wave experiences a refraction effect towards the magnetic null point.  The magnitude of the refraction is different for fluid elements approaching the null from various directions and is governed by the  Alfv\'en speed profile, $v_A^2={x^2+\epsilon^2y^2+\left(\epsilon+1\right)^2z^2}$ (in non-dimensionalised variables) and it is this different dependence on $x$, $y$ and $z$  that lead to  different strength refraction effects. However, for all $\epsilon$ the main result holds: the fast wave accumulates at the null point.

In both Sections \ref {sub1} and \ref{sub2}, the fast wave, and thus the wave energy, accumulates at the null point. The fast wave cannot cross the null because the Alfv\'en speed there is zero. Thus, the length scales between the leading and trailing edges of wave pulses will decrease  indicating that the current (and all other gradients) will increase. In Section \ref{current_fast}, we calculated theoretical limits of the current build up and found that it was bounded by two exponentially growing functions. Moreover, it was shown that for a fast wave starting at $z=z_0$, $\left|{\bf{j}}_1\right|\sim z^{-1} \sim  {\rm{e}}^{\left( \epsilon +1 \right) t}$, and for fast wave starting at $y=y_0$: $\left|{\bf{j}}_1\right| \sim y^{-1} \sim  {\rm{e}}^{ \epsilon t}$. Hence, no matter how small the value of the resistivity is, if we include the dissipative term then eventually the $\eta \nabla ^2 {{\bf{B}}}_1$ term in Equation (\ref{6}) will become non-negligible and dissipation will become important. In addition, since ${\bf{j}}_1$ grows exponentially in time, diffusion terms will become important in a time ${\sim}\:{\log{\eta}}$; as found by \inlinecite{CW1992} and \inlinecite{CM1993}. This means that linear wave dissipation will be very efficient. Thus, we deduce that 3D null points will be the locations of wave energy deposition and preferential heating.

We find that the Alfv\'en wave propagates along the fieldlines, and that an Alfv\'en wave fluid element is confined to the fieldline it starts on. For the Alfv\'en wave approaching the null point from above (planar wave starting at $z=z_0$) the wave accumulates along the fan plane. For an Alfv\'en wave approaching from the side (propagation initially perpendicular to the spine) the wave accumulates along the spine. This behaviour is in good agreement with the results of  \inlinecite{PG2007} and \inlinecite{PBG2007}, but the method we present here clearly illustrate why this occurs, e.g. by following the ray paths in Section \ref{sub4}, it is clear why an Alfv\'en wave generated crossing the fan plane must accumulate along the spine.

Furthermore, we found an analytical solution for the Alfv\'en wave. From this we were able to show that the Alfv\'en wave rotates the fieldlines, the volume occupied by the wave  pulse is conserved and that the associated currents  build up  exponentially. For a wave starting at $z=z_0$, the currents build up along the fan plane, and  $j_x$ and $j_y$ grow as ${{\rm{e}}}^{\left( \epsilon +1 \right) t}$ and  ${{\rm{e}}}^{2\epsilon t}$, respectively. Thus,  resistive effects will eventually become non-negligible in a time $\sim \log{\eta}$ . For a wave starting at $z=z_0$, the value of $\epsilon$ determines where the preferential heating will occur:  fan-plane ($\epsilon=1$), along the $x-$axis ($0<\epsilon <1$) or along the $y-$axis ($\epsilon>1$). In contrast, an Alfv\'en wave starting at $x=x_0$ or $y=y_0$ will lead to  preferential heating along the spine.

All of the work described here highlights the importance of understanding the magnetic topology of a system, specifically the location of the spines and fans for a 3D null point. It is at these areas where preferential heating will occur, {\it{i.e.}} these areas are where the wave energy accumulates. In addition, it is of note that for both the fast and Alfv\'en waves, current builds up exponentially and thus  {\emph{diffusion terms will become important in a time that depends on $\log { \eta }$}}. This is all in good agreement with the 2D work of McLaughlin \& Hood (\citeyear{MH2004}; \citeyear{MH2005}; \citeyear{MH2006}).

It is also useful to make an order of magnitude estimate for the quantities presented here, in order to gain a better understanding of the physical conclusions. Let us consider  our $\epsilon=1$ system to have characteristic length $L=10$\,Mm, $B=10$\,G, $\rho_0= 10^{-12}$kg$\:$m$^{-3}$, $\mu=4\pi\times 10^{-7}$H$\:$m$^{-1}$ and $\eta=1$\,m$^2$$\:$s$^{-1}$. This gives a characteristic speed of ${\bar{v}}= 892$\,km s$^{-1}$, a characteristic time ${\bar{t}}=L / {\bar{v}} = 11.2$\,seconds, frequency $\omega_0= 2\pi/ {\bar{t}}=0.56$\,Hz, wavelength $\lambda= {\bar{v}} / \omega_0=  1.59$\,Mm and ${\bar{j}} = B/\mu L = 8\times 10^{-5}$A. Thus, for the planar fast wave starting at $z_0=L$ and considering the behaviour along $x=y=0$, we find that after a time $t=1$\,second, we have built up a current of $0.3$\,mA  (Equation \ref{example}). We can also estimate the time it takes for resistive effects to become important. We assume ${\partial {\bf{B}}}/{\partial t} \approx \eta \nabla^2 {\bf{B}}\Rightarrow \omega B \approx \eta B / (\delta z)^2$, where  $\delta z$ is the distance between the leading edge and trailing edge of our wave pulse, and we take  $\delta z=\lambda$, where the form of $z$ is given by Equation ({\ref{chris}}). We find that resistive effects become non-negligible in a time $t\approx {\log{\left(\omega \lambda^2\right / \eta)}}/4=7$\,seconds. For comparison, after $t=7$\,seconds, our wave has built up a current of $0.87$\,mA and has travelled a distance of $7.13$\,Mm. The Alfv\'en wave is degenerate with the fast wave along  the spine and so has the same estimates as above (under identical conditions).

The 3D WKB technique  described in this project can also be easily applied to other magnetic configurations, e.g.  3D dipole, and we hope that this paper has illustrated the potential of the technique. In addition, it is possible to extend the work by dropping the cold plasma assumption. This will lead to a third root of Equation (\ref{F1}) which will correspond to the behaviour of the slow magnetoacoustic wave.

We conclude this paper with some caveats concerning the method presented here, {\it{i.e.}} if modellers wish to compare their work with a WKB approximation, it is essentialy to know the limitations of such a method.  Firstly, in linear 3D MHD, we would expect a coupling between the fast and Alfv\'en wave types due to the geometry. However, under the WKB approximation presented here, the wave sees the field as locally uniform  and so there is no coupling between the wave types. To include the coupling, one needs to include the next terms in the approximation, {\it{i.e.}} the work presented here only deals with the first-order terms of the WKB approximation.

Secondly, note that the  work here is only strictly valid for high-frequency waves, since we took $\phi$ and hence $\omega=\phi_t$ to be a large parameter in the system. The extension to low frequency waves is considered in \inlinecite{Weinberg1962}.

Finally, the WKB approximation becomes degenerate at the points $v_A=c_s$, {\it{i.e.}} regions where the Alfv\'en speed and sound speed are equal.  Thus, the WKB method in the form presented here cannot be used to investigate mode conversion ({\it{e.g.}} see \opencite{MH2006b}) and, as mentioned above, the next terms in the approximation are needed. Alternatively, work is underway to  overcome this degeneracy using the method developed by \inlinecite{Cairns} to match WKB solutions across the mode conversion layer (layer where $v_A=c_s$). The results of such work in 1D can be found in \inlinecite{Dee}.



\appendix

\section{${\bf {k}}$ parallel to ${\bf B}_{0}$}\label{appendixA}

In this appendix, we address the scenario $   {\bf {k}} = \lambda       {\bf B}_{0}          $ in which the vectors of our three-dimensional coordinate system   $( {\bf B}_{0},  {\bf k},  {\bf B}_{0} \times {\bf k}$) are no longer linearly independent.  To do this we consider the following Equation:  
\begin{eqnarray}
\frac{\partial^2 } {\partial t^2}{\mathbf{v}}_1 = c_s^2 \nabla \left( \nabla \cdot {\mathbf{v}}_1 \right) + \left\{ \nabla \times \left[  \nabla \times \left( {\mathbf{v}}_1 \times {\mathbf{B}}_0 \right) \right] \right\} \times {\mathbf{B}}_0   \;\;, \label{1000}
\end{eqnarray}
which is derived in the same way as Equation (\ref{10}) but without assuming a cold plasma.  Under $c_s=0$, Equation (\ref{1000}) reduces to Equation (\ref{10}). Thus, assuming  ${\bf {k}} = \lambda       {\bf B}_{0}$ and applying the WKB approximation (Equation \ref{WKB}) to Equation (\ref{1000}) gives:
\begin{eqnarray*}
\omega ^2 {\bf{v}} &=& c_s^2  \left( {\bf{k}} \cdot {\bf{v}} \right) {\bf{k}} +   \left\{  {\bf{k}}  \times \left[   {\bf{k}}  \times \left( {\bf{v}}\times {\mathbf{B}}_0 \right) \right] \right\} \times \frac{{\mathbf{B}}_0 }{\mu \rho_0} \\
&=&   c_s^2  \left( {\bf{k}} \cdot {\bf{v}} \right) {\bf{k}} + \left( {\bf{k}} \cdot {\bf{B}}_0 \right)^2 \frac{{\bf{v}}}{\mu \rho_0}  -   \left( {\bf{k}} \cdot {\bf{B}}_0 \right)   \left( {\bf{v}} \cdot {\bf{B}}_0 \right)\frac{{\bf{k}}}{\mu \rho_0}\\
 &-&  \left( {\bf{k}} \cdot {\bf{B}}_0 \right)   \left( {\bf{k}} \cdot {\bf{v}} \right)\frac{{\bf{B}}_0}{\mu \rho_0} +  \left( {\bf{k}} \cdot {\bf{v}} \right) \left| {\bf{B}}_0 \right|^2 \frac{{\bf{k}}}{\mu \rho_0} \\
&=&  c_s^2 \lambda^2 \left( {\bf{B}}_0 \cdot {\bf{v}} \right) {\bf{B}}_0 + \lambda^2   \frac{\left| {\bf{B}}_0 \right|^2}{\mu \rho_0} \left| {\bf{B}}_0 \right|^2 {\bf{v}}   -  \lambda^2  \frac{\left| {\bf{B}}_0 \right|^2}{\mu \rho_0}   \left( {\bf{v}} \cdot {\bf{B}}_0 \right){\bf{B}}_0 \\
&-&  \lambda^2   \frac{ \left| {\bf{B}}_0 \right|^2}{\mu \rho_0} \left( {\bf{B}}_0 \cdot {\bf{v}} \right){\bf{B}}_0 + \lambda^2  \left( {\bf{B}}_0 \cdot {\bf{v}} \right) \frac{\left| {\bf{B}}_0 \right|^2}{\mu \rho_0} {\bf{B}}_0 \\
&=&  c_s^2 \lambda^2 \left( {\bf{B}}_0 \cdot {\bf{v}} \right) {\bf{B}}_0 +  \lambda^2   v_A^2 \left| {\bf{B}}_0 \right|^2 {\bf{v}}   -  \lambda^2  v_A^2  \left( {\bf{v}} \cdot {\bf{B}}_0 \right){\bf{B}}_0 
\end{eqnarray*}
where  $v_A^2=\frac{\left| {\bf{B}}_0 \right|^2}{\mu \rho_0}$ and we have explicitly included $\mu$ and $\rho_0$. Thus, for  ${\bf{v}}$ parallel to ${\bf{{B}}_0}$, {\it{i.e.}} ${\bf{v}} =\alpha {\bf{{B}}_0}$, we have:
\begin{eqnarray*}
\omega ^2 \alpha {\bf{B}}_0&=& c_s^2 \lambda^2 \alpha \left| {\bf{B}}_0  \right|^2 {\bf{B}}_0 + \lambda^2   v_A^2 \left| {\bf{B}}_0 \right|^2 \alpha {\bf{B}}_0   -  \lambda^2  v_A^2 \alpha \left|{\bf{B}}_0 \right|^2 {\bf{B}}_0\\  \Rightarrow \quad\omega ^2  &=&c_s^2 \left| {\bf{k}}  \right|^2  
\end{eqnarray*}
So the longitudinal oscillations  (since ${\bf{v}}\parallel {\bf{B}_0} \parallel {\bf{k}}$) propagate at the sound speed, {\it{i.e.}} this is the dispersion relation for slow waves.

For ${\bf{v}}$ perpendicular to ${\bf{B}_0}$ $\left({\bf{v}} \cdot {\bf{B}_0}=0\right)$, {\it{i.e.}} transverse oscillations, we have:
\begin{eqnarray*}
\omega ^2 {\bf{v}}_\perp = \lambda^2   v_A^2 \left| {\bf{B}}_0 \right|^2 {\bf{v}}_\perp \quad \Rightarrow \quad \omega ^2 =    v_A^2 \left| {\bf{k}} \right|^2 
\end{eqnarray*}
This is the dispersion relation for a transverse and  incompressional Alfv\'en wave ({\it{i.e.}}  ${\bf{k}}  \parallel {\bf{B}}_0 \perp \bf{v}$). However, it is also the dispersion relation for the fast magnetoacoustic wave propagating in the direction of the magnetic field. Thus, we cannot distinguish between these two wave types in this specific scenario.

It is also worth noting that even though the coordinate system we considered in Section \ref{WKBAPPROXIMATION} is not linearly independent when ${\bf{B}_0} \parallel {\bf{k}}$, the result, Equation (\ref{F1}), still holds. Under the assumption  $   {\bf {k}} = \lambda       {\bf B}_{0}          $, Equation (\ref{F1}) simplifies to:
\begin{eqnarray*}
 \mathcal{F}\left( \phi, x,y,z,p,q,r \right)&=&  \left( \omega  - v_A^2      \left|   {\bf{k}} \right|^2  \right)^2     =0\\
\end{eqnarray*}
So we have a double root and the solution is degenerate, {\it{i.e.}} it is impossible to distinguish the waves under these conditions (in agreement with the work above).

\section{($\epsilon=1/2$) Planar fast wave starting at $y_0=1$ and $x_0=1$}\label{extra}

\begin{center}
\begin{figure}[t]
\includegraphics[width=4.8in]{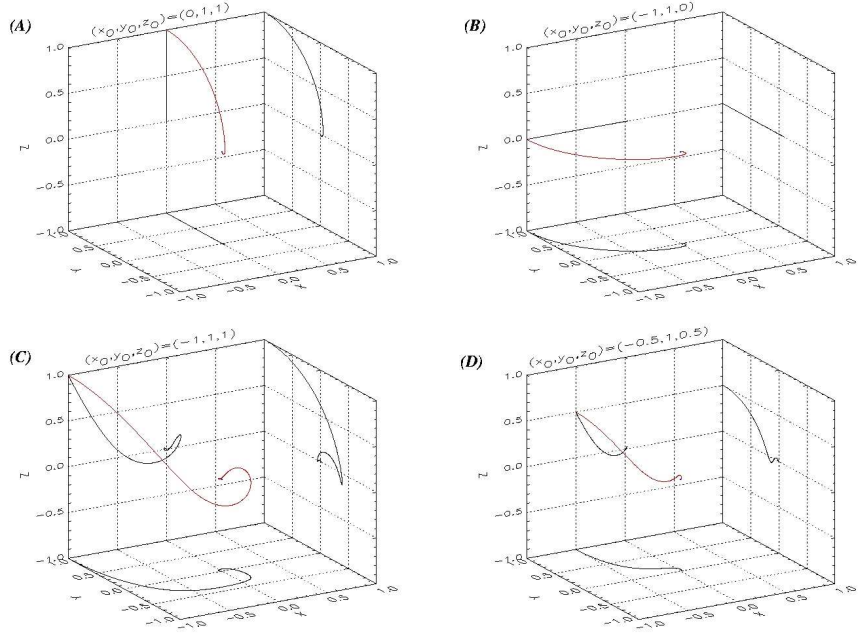}
\caption{($\epsilon=1/2$) Ray paths for fluid elements that begin at $(x_0,y_0,z_0)=$ $(A)$ $(0,1,1)$, $(B)$  $(-1,1,0)$, $(C)$ $(-1,1,1)$ and $(D)$ $(-0.5,1,0.5)$.  This figure is also available as an mpg animation showing many more starting points. Red indicates the 3D ray path and black indicates the $xy$, $yz$ and $xz$ projections of this ray path onto the respective planes.  The blue dot indicates the position of the magnetic null point.}
\label{extra1}
\end{figure}
\end{center}

\begin{center}
\begin{figure}[t]
\includegraphics[width=4.8in]{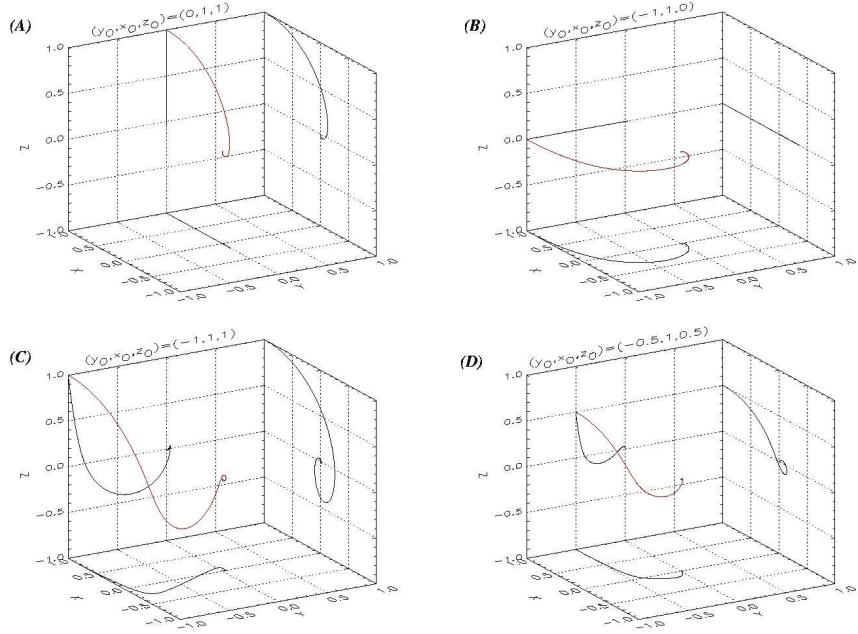}
\caption{($\epsilon=1/2$) Ray paths for fluid elements corresponding to a fast wave starting at $x_0=1$. We have made the transformation $(x,y)\rightarrow (-y,x)$ in order to more easily compare with Figure \ref{extra1}. Thus, this figure shows ray paths that begin at $(y_0,x_0,z_0)=$ $(A)$ $(0,1,1)$, $(B)$  $(-1,1,0)$, $(C)$ $(-1,1,1)$ and $(D)$ $(-0.5,1,0.5)$. This figure is also available as an mpg animation in the electronic edition of {\it{Solar Physics}}, showing many more starting points. Red indicates the 3D ray path and black indicates the $xy$, $yz$ and $xz$ projections of this ray path onto the respective planes.  The blue dot indicates the position of the magnetic null point.}
\label{extra2}
\end{figure}
\end{center}

 The ray paths corresponding to a planar fast wave starting at $y_0=1$ and starting at $x_0=1$ in the $\epsilon=1/2$ magnetic configuration can be found in Figures \ref{extra1} and \ref{extra2}. Recall that the fast wave cannot cross the null because the Alfv\'en speed there is zero.

\section{Limits on fast wave current build up}\label{limits}


Define  $R^2=  x^2+y^2+z^2$ and assume  $0\le \epsilon \le 1$. Recall $\left| {\bf{B}}_0\right| ^2= x^2+\epsilon^2 y^2 + \left( \epsilon +1 \right)^2 z^2$. This can be bounded above by $\left(\epsilon +1 \right)^2 R^2$, since $1 \le \left(\epsilon +1 \right)^2$ and  $\epsilon^2 \le \left(\epsilon +1 \right)^2$. Hence, using a similar lower bound, we have:
\begin{eqnarray}
\epsilon^2 R^2 \le \left| {\bf{B}}_0\right| ^2 \le \left(\epsilon +1 \right)^2 R^2 \;\;, \label{ant}
\end{eqnarray}
These limits can also be understood physically:  Recall that constant values of $\left| {\bf{B}}_0\right| ^2$ defines an ellipsoid. Since we assume $0\le \epsilon \le 1$, the largest distance from the centre to any edge of the ellipsoid is $\left(\epsilon +1\right) R$, and the smallest distance  is  $\epsilon R$. Thus, physically we have encased our ellipsoid inside two spheres of radii  $\epsilon R$ and $\left(\epsilon +1\right) R$.

From Equation (\ref{dalembertfast}) we have:
\begin{eqnarray*}
\frac{{\rm{d}}}{{\rm{d}}s} R^2 &=& 2x \frac{{\rm{d}}x}{{\rm{d}}s} +2y \frac{{\rm{d}}y}{{\rm{d}}s} +2z \frac{{\rm{d}}z}{{\rm{d}}s}\\
 &=& -2\left( xp+yq+zr\right) \left| {\bf{B}}_0\right| ^2 = -2 \left(x_0p_0+y_0q_0+z_0r_0\right)  \left| {\bf{B}}_0\right| ^2 \;\;,
\end{eqnarray*}
where we have used the conserved quantity from Equation (\ref{bbbbbb}). Define $\alpha=x_0p_0+y_0q_0+z_0r_0$. Thus, from Equation (\ref{ant}) we have the inequality:
\begin{eqnarray*}
  -2 \alpha \left(\epsilon+1\right)^2 R^2         \le  \frac{{\rm{d}}}{{\rm{d}}s} R^2  \le -2 \alpha  \epsilon^2 R^2  \;\;,
\end{eqnarray*}
We can integrate and invert this inequality to obtain:
\begin{eqnarray}
 R_0^2 {\rm{e}}^{ -2 \alpha \left(\epsilon+1\right)^2 s}  \le R^2 \le  R_0^2 {\rm{e}}^{ -2 \alpha \epsilon^2 s} \quad \Rightarrow  \quad \frac{1}{R_0^2} {\rm{e}}^{ 2 \alpha \epsilon^2 s} \le \frac{1}{R^2} \le \frac{1}{R_0^2}  {\rm{e}}^{ 2 \alpha \left(\epsilon+1\right)^2 s}\label{dec}
\end{eqnarray}
where $R_0$ is a constant that depends upon starting position: $R_0^2=x_0^2+y_0^2+z_0^2$. Hence, inverting Equation (\ref{ant}) and combining it with  Equation (\ref{dec}) gives:
\begin{eqnarray*}
\frac{1}{  { \left(\epsilon+1\right)^2}       R_0^2 }  {\rm{e}}^{ 2 \alpha \epsilon^2 s} \le \frac{1}{\left(\epsilon+1\right)^2 R^2} \le \frac{1}{\left| {\bf{B}}_0\right| ^2} \le \frac{1}{\epsilon^2 R^2} \le   \frac{1}{  \epsilon^2      R_0^2}{\rm{e}}^{ 2 \alpha \left(\epsilon+1\right)^2 s} \;\;.
\end{eqnarray*}
Finally, we recall $t=\omega_0 s$ and thus:
\begin{eqnarray}
\frac{1}{  { \left(\epsilon+1\right)^2}       R_0^2 }  {\rm{e}}^{ 2 \alpha \epsilon^2 t / \omega_0} \le \frac{1}{\left| {\bf{B}}_0\right| ^2} \le  \frac{1}{  \epsilon^2      R_0^2}{\rm{e}}^{ 2 \alpha \left(\epsilon+1\right)^2 t / \omega_0} \;\;.\label{kaj}
\end{eqnarray}

\section{Volume}\label{Volume}


Assume we generate an initially rectangular wave pulse of volume $V_0= \left( x_1-x_2\right) \times \left( y_1-y_2\right)\times \left( z_1-z_2\right)$, where $x_1$, $x_2$, $y_1$, $y_2$, $z_1$ and $z_2$ define the starting points at the edges of our domain. The wave will evolve according to Equation (\ref{hhhh}) and thus, after travelling distance $s$ along the characteristic curve, will occupy  a volume:
\begin{eqnarray*}
V_{end}&=&\left( x_1{\rm{e}}^{\omega s}-x_2{\rm{e}}^{\omega s}\right) \times \left( y_1{\rm{e}}^{ \epsilon  \omega s}-y_2{\rm{e}}^{  \epsilon  \omega s}\right)\times \left( z_1{\rm{e}}^{- \left( \epsilon +1 \right) \omega s}-z_2{\rm{e}}^{- \left( \epsilon +1 \right) \omega s}\right)\\
&=&\left( x_1-x_2\right) \times \left( y_1-y_2\right)\times \left( z_1-z_2\right)=V_0
\end{eqnarray*}
Thus, volume is conserved  for an Alfv\'en wave in this system.


\begin{acks}
JSLF acknowledges financial assistance from a Cormack Vacation Research Scholarship awarded by the Royal Society of Edinburgh. JAM wishes to thank the Royal Astronomical Society for awarding him a RAS grant to travel to the SOHO19 conference (where this work was first presented). JAM also acknowledges financial assistance from the St Andrews STFC Rolling Grant and from the Leverhulme Trust. JAM wishes to thank  Jesse Andries, Ineke De Moortel, and Jaume Terradas  for insightful discussions. AWH and JAM also wish to thank Clare Parnell for helpful     suggestions regarding this paper.
\end{acks}

\end{article} 

\end{document}